\documentstyle{article}
\long\def\killtext#1{}

  \title{Information Distance\thanks{These results were announced 
in {\em
Proc. 25th ACM Symp. Theory of Comput.}, 1993, 21-30.}}

 \author{  Charles H.~Bennett\thanks{
 T.J.\ Watson\ IBM\ Research Cen\-ter, 
 York\-town Heights, NY 10598, USA.
 Email: bennetc@watson.ibm.com.}
 \and 
  P\'eter G\'acs\thanks{
 Computer\ Sci.\ Dept.,
 Boston University,
 Boston, MA 02215 USA.
 Email: gacs@cs.bu.edu.
  Part of this research was done during the author's stay at IBM
Watson Research Center.
  Partially supported by NSF grant CCR-9002614, and by NWO through NFI
Project ALADDIN under Contract number NF 62-376 and Scientific Visitor
Award B 62-394.}
  \and
 Ming Li\thanks{
 Computer Sci.\ Dept., 
 University of Waterloo,
 Waterloo, Ontario, N2L 3G1 Canada.
 Email: mli@math.uwaterloo.ca. 
  Partially supported by NSERC Operating  grant OGP-046506.}
  \end{tabular}
  \vskip 1em 
  \begin{tabular}[t]{c}%
 Paul M.B.~Vit\'anyi\thanks{
 CWI and University of Amsterdam.  Address:
 CWI, Kruislaan 413
 1098 SJ Amsterdam, The Netherlands.
 Email: paulv@cwi.nl.
 Partially supported by NSERC International Scientific Exchange Award
ISE0046203, 
by the European Union
through NeuroCOLT ESPRIT Working Group Nr. 8556,
and by  NWO through NFI Project ALADDIN under Contract
number NF 62-376.}
 \and 
  Wojciech H.~Zurek\thanks{
 Theor.\ Div., Los Alamos National Laboratories and Santa F\'e Inst.
  Address: Los Alamos, NM 87545, USA.
 Email: whz@lanl.gov}
}

 \makeatletter

\def\thanks#1{\footnotemark\begingroup
\def\protect{\noexpand\protect\noexpand}\xdef\@thanks{\@thanks
  \protect\footnotetext[\the\c@footnote]{#1}}\endgroup}
\def\@thanks{}
\def\and{
\end{tabular}\hskip 1em plus .17fil\begin{tabular}[t]{c}
}

\def\maketitle{\par
 \begingroup
 \def\thefootnote{\fnsymbol{footnote}}
 \def\@makefnmark{\hbox 
 to 0pt{$^{\@thefnmark}$\hss}} 
 \if@twocolumn 
 \twocolumn[\@maketitle] 
 \else \newpage
 \global\@topnum\z@ \@maketitle \fi\thispagestyle{plain}\@thanks
 \endgroup
 \setcounter{footnote}{0}
 \let\maketitle\relax
 \let\@maketitle\relax
 \gdef\@thanks{}\gdef\@author{}\gdef\@title{}\let\thanks\relax}

\makeatletter

\long\def\killtext#1{}
\newenvironment{formal}[1]{ \refstepcounter{equation}
 \trivlist\item[\hskip \labelsep {\bf
  (\csname theequation\endcsname)\ #1\hskip\labelsep}]}{\hspace*{\fill}$\Diamond$\endtrivlist}

\newenvironment{formalsl}[1]{ \refstepcounter{equation}
 \trivlist\item[\hskip \labelsep {\bf
  (\csname theequation\endcsname)\ #1\hskip\labelsep}]\sl}{\endtrivlist}

 \@addtoreset{equation}{section}

 \def\thetable{\@arabic\c@table}

 \newcounter{proofpart}
 \def\proofpart{\refstepcounter{proofpart}\@ifnextchar[{\@proofpart
   }{\@proofpartu}}
 \def\@proofpartu{\paragraph{\sf \arabic{proofpart}.}}
 \def\@proofpart[#1]{\paragraph{\sf
   \arabic{proofpart}.\hskip\labelsep #1}}

\newenvironment{proof}{\par \bf Proof.\rm}{\hspace*{\fill}$\Box$\vspace{1ex}}

 \newcommand\setof[1]{\mathopen\{\,#1\,\mathclose\}}
 
 \newcommand\ngl[1]{{\langle #1\rangle}}

 \def\ol#1{#1^*}

 \renewcommand \lg{\lambda} 

\makeatother

\newcommand{\lea}{\stackrel{+}{<}}
\newcommand{\gea}{\stackrel{+}{>}}
\newcommand{\eqa}{\stackrel{+}{=}}

\newcommand{\lel}{\stackrel{\log}{<}}
\newcommand{\gel}{\stackrel{\log}{>}}
\newcommand{\eql}{\stackrel{\log}{=}}
\newcommand{\UR}{\mbox{\it UR}}
\newcommand{\KR}{\mbox{\it KR}}

\makeatother
\begin{document}
\maketitle

\begin{abstract}
  While Kolmogorov complexity is the accepted absolute measure of information
content in an individual finite object, a
similarly absolute notion is needed for the information
distance between two individual objects, for example, two pictures.
  We give several natural definitions of a universal
information
metric, based on length of shortest programs for either ordinary
computations or reversible
(dissipationless) computations.
  It turns out that these definitions are equivalent up to an additive
logarithmic term.
  We show that the information distance is 
a universal 
cognitive similarity distance.
  We investigate the maximal correlation of the shortest programs
involved, the maximal uncorrelation of programs (a generalization of the
Slepian-Wolf theorem of classical information theory), and the density
properties of the discrete metric spaces induced by the information
distances.
  A related distance measures the amount of nonreversibility of a
computation.
  Using the physical theory of reversible computation, we 
give an appropriate (universal, anti-symmetric, and transitive)
measure of the thermodynamic work required to transform
one object in another object by the most efficient process. 
 Information distance between individual objects 
is needed
in pattern recognition where one wants to express
effective notions of ``pattern similarity'' or ``cognitive similarity''
between individual objects
and in thermodynamics of computation where one wants to analyse
the energy dissipation of a
computation from a particular input to a particular output.


  \end{abstract}

\newpage

 \section{Introduction}\label{sect.intro}
  We write {\em string} to mean a finite binary string.
  Other finite objects can be encoded into strings in natural
ways.
  The set of strings is denoted by $\{0,1\}^*$.

  The Kolmogorov complexity, or algorithmic entropy, $K(x)$ of a
string $x$ is the length of a shortest binary program to compute
$x$ on a universal computer (such as a universal Turing machine). 
  Intuitively, $K(x)$ represents the minimal amount of information
required to generate $x$ by any effective process, \cite{Ko65}.
  The conditional Kolmogorov complexity $K(x|y)$ of $x$ relative to
$y$ is defined similarly as the length of a shortest program
to compute $x$ if $y$ is furnished as an auxiliary input to the
computation.
  The functions $K( \cdot)$ and $K( \cdot| \cdot)$, 
though defined in terms of a
particular machine model, are machine-independent up to an additive
constant
 and acquire an asymptotically universal and absolute character
through Church's thesis, from the ability of universal machines to
simulate one another and execute any effective process.
  The Kolmogorov complexity of a string can be viewed as an absolute
and objective quantification of the amount of information in it.
   This leads to a theory of {\em absolute} information {\em contents}
of {\em individual} objects in contrast to classical information theory
which deals with {\em average} information {\em to communicate}
objects produced by a {\em random source}.
  Since the former theory is much more precise, it is surprising that
analogons of theorems in classical information theory hold for
Kolmogorov complexity, be it in somewhat weaker form.

  Here our goal is to study the question of an
``absolute information distance metric''
between individual objects. This should be contrasted with an information
metric (entropy metric)
such as $H(X|Y)+H(Y|X)$
 between stochastic sources $X$ and $Y$.
Non-absolute approaches to information distance between individual 
objects have been
studied in a statistical setting, 
see for example \cite{ZM93} for a notion of empirical
information divergence (relative entropy) between two
individual sequences. Other approaches include various
types of edit-distances between pairs of strings: the minimal number
of edit operations from a fixed set required to transform one string in the 
other string. Similar distances are defined on trees or other data structures.
The huge literature on this ranges from pattern matching and cognition
to search strategies on internet and computational
biology. As an example we mention nearest neighbor interchange distance
between evolutionary trees in computational biology,
\cite{STT92,LTZ96}.
  {\em A priori} it is not immediate what is the most appropriate
{\em universal} symmetric
informational distance between two strings, that is, the minimal
quantity of information sufficient to translate between $x$ and $y$,
generating either string effectively from the other.
  We give evidence that such 
 notions are relevant for pattern recognition, cognitive sciences
in general, various application areas, 
 and physics of
computation.

\paragraph{Metric.} A {\em distance} function $D$ with nonnegative
real values, defined on the Cartesian product $X \times X$ of
a set $X$ is called a {\em metric}
on $X$ if for every $x,y,z \in X$:
\begin{itemize}
\item
$D(x,y)=0$ iff $x=y$ (the identity axiom);
\item
$D(x,y)+D(y,z) \geq D(x,z)$ (the triangle inequality);
\item
$D(x,y)=D(y,x)$ (the symmetry axiom).
\end{itemize}
A set $X$ provided with a metric is called a {\em metric space}.
For example, every set $X$ has the trivial {\em discrete metric}
$D(x,y)=0$ if $x=y$ and $D(x,y)=1$ otherwise. All information
distances in this paper are defined on the set $X=\{0,1\}^*$
and satisfy
the metric conditions up to an additive
constant or logarithmic term while the identity axiom can be
obtained by normalizing.

\paragraph{Algorithmic Information Distance.}
  Define the information distance
as the length of a shortest binary program that
computes $x$ from $y$ as well as computing $y$ from $x$.
  Being shortest, such a program should take advantage of any
redundancy between the information required to go from $x$ to $y$
and the information required to go from $y$ to $x$.
  The program functions in a catalytic capacity in the
sense that it is required to transform the
input into the output, but itself remains
present and unchanged throughout the computation.
  We would like to know to what extent the information
required to compute $y$ from $x$ can be made to overlap with that
required to compute $x$ from $y$.
  In some simple cases, {\em complete} overlap can be achieved, so that
the same minimal program suffices to compute $x$ from $y$ as to
compute $y$ from $x$.
  For example if $x$ and $y$ are independent random binary strings of
the same length $n$ (up to additive contants $K(x|y)=K(y|x)=n$), then
their bitwise exclusive-or $x \oplus y$ serves as a minimal program
for both computations.
  Similarly, if $x=uv$ and $y=vw$ where $u$, $v$, and $w$ are
independent random strings of the same length, then $u \oplus w$ plus
a way to distinguish $x$ from $y$ is a
minimal program to compute either string from the other.

\paragraph{Maximal Correlation.}
  Now suppose that more information is required for one of these
computations than for the other, say,
 \[
  K(y|x) > K(x|y).
 \]
  Then the minimal programs cannot be made identical because they
must be of different sizes.
  In some cases it is easy to see that the overlap 
can still be made complete, in
the sense that the larger program (for $y$ given $x$) can be made to
contain all the information in the shorter program, as well as some
additional information.
  This is so when $x$ and $y$ are independent random strings of
unequal length, for example $u$ and $vw$ above.
  Then $u\oplus v$ serves as a minimal program for $u$ from $vw$, and
$(u \oplus v)w$ serves as one for $vw$ from $u$.

 A principal result of this paper in Section~\ref{s.conv} shows that,
up to an additive  logarithmic error term, 
the information required to translate
between two strings can be represented in this maximally overlapping
way in every case.
  Namely, let
  \begin{eqnarray*}
    && k_1 = K(x|y),\ k_2 = K(y|x),
\\  && l   = k_2-k_1
 \end{eqnarray*}
  where we assume $k_1\le k_2$.
  Then there is a string $q$ of length $k_1+ K(k_1,k_2)$ and a string
$d$ of length $l$ such that $q$ serves as the minimal
program both to compute from $xd$ to $y$ and from $y$ to $xd$.
The term $ K(k_1,k_2)$ has magnitude $ O (\log k_2 )$.
  This means that the information to pass from $x$ to $y$ can
always be maximally correlated with the information to get
from $y$ to $x$.
  It is therefore never the case that a large amount of
information is required to get from $x$ to $y$ and a large {\em
but independent} amount of information is required to get from
$y$ to $x$.
  This demonstrates that
 \[
  E_1(x,y)= \max\{K(y|x),K(x|y)\}
 \]
  equals the length of a shortest program $p:=qd$ 
to compute $x$
from $y$ and $y$ from $x$, up to a logarithmic additive term.\footnote
 {The situation is analogous to the inverse function theorem
of multidimensional analysis.
  This theorem says that under certain conditions, if we have a vector
function $f(x,p)$ then it has an inverse $g(y,p)$ such that in a
certain domain, $f(x,p)=y$ holds if and only if $g(y,p)=x$.
  In the function going from $y$ to $x$, the parameter $p$ remains the
same as in the function going from $x$ to $y$.}
  (It is very important here that the time of computation is
completely ignored: this is why this result does not contradict the
idea of one-way functions.)

  The process of going from $x$ to $y$ may be broken into two stages.
  First, add the string $d$; second, use the difference program $q$
between $xd$ and $y$.
  In the reverse direction, first use $q$ to go from $y$ to $xd$;
second, erase $d$.
  Thus the computation from $x$ to $y$ needs both $q$ and $d$, that is,
the program $p=qd$, while
the computation from $y$ to $x$ needs only $q$ as program.

\paragraph{Minimal Correlation.}
  The converse of maximal correlation is that in the special case of
the shortest programs for going between {\em independent random} 
$x$ and $y$, they
can be choosen {\em completely independent}.
  For example use $y$ to go from $x$ to $y$ and $x$ to go from $y$ to
$x$.
  This turns out to hold also in the general case
 for arbitrary pairs $x,y$, as will be shown
in Theorem \ref{t.Slepian-Wolf}, but only with respect to an
``oracle'': a certain constant string that must be in all the
conditions.
  This theorem can be considered a generalization of the Slepian-Wolf
Theorem of classical information theory~\cite{CsiszarKorner80}.

\paragraph{Universal Cognitive Distance.}
  Section \ref{s.axioms} develops an axiomatic theory of ``pattern
distance'' or more generally a ``cognitive similarity metric''
 and argues that the function $E_1(x,y)$ is the most natural
way of formalizing a {\em universal} cognitive
distance between $x$ and $y$.
  This nonnegative function is $0$ iff $x=y$ (rather, its normalized
  version in Theorem~\ref{t.optimal.cognitive.dist} satifies this),
it is symmetric, obeys the triangle inequality to within an
additive constant, and is minimal among the class of
distance functions that are computable in a weak sense
and satisfy
a normalization constraint limiting the number of distinct
strings $y$ within a given distance of any $x$.
It uncovers all effective similarities between two individual objects.

\paragraph{Information Distance for Reversible Computation.}
  Up till now we have considered
ordinary computations, but if one insists
that the computation be performed {\em reversibly\/}, that is by a
machine whose transition function is
one-to-one~\cite{Lecerf63,Bennett73}, then the full program $p=qd$ above is
needed to perform the computation in either direction.
  This is because reversible computers cannot get rid of unwanted
information simply by erasing it as ordinary irreversible computers
do.
  If they are to get rid of unwanted information at all, they must
cancel it against equivalent information already present elsewhere in
the computer. Reversible computations are discussed in Section
\ref{s.rev} where we
  define a reversible distance
 $E_2(x,y)= \KR(x|y)= \KR(y|x)$, representing the amount of information
required to program a reversible computation from $x$ to $y$ (which by
definition is the reverse of the computation from $y$ to $x$).
  The $E_2$ distance is equal within an additive constant to the
length of the conversion program $p=qd$ considered above, and so is at
most greater by an additive logarithmic term than the optimal distance $E_1$.
It is also a metric.
  The reversible program functions again in a catalytic manner.

  Hence, three very different definitions arising from different
backgrounds identify up to logarithmic additive terms the same notion of
information distance and corresponding metric. It is compelling to believe
that our intuitive notions are adequately formalized by
this universal and absolute notion of information metric.

\paragraph{Minimal Number of Irreversible Operations.}
  Section \ref{s.alt-dist} considers reversible computations
where the program is not catalytic but
in which additional information $p$ (like a program) besides $x$ is
consumed, and additional information $q$ (like garbage) besides $y$ is
generated and irreversibly erased.
  The sum of these amounts of information, defined as distance
$E_3(x,y)$, represents the minimal number of irreversible bit
operations in an otherwise reversible computation from 
$x$ to $y$ in which the program is
not retained.
  It is shown to be equal to within a logarithmic term to Zurek's sum
metric $K(y|x) + K(x|y)$, which is typically larger than our proposed
optimal metric $E_1$ because of the redundancy between $p$ and $q$.
  But using the program involved in $E_1$ we both consume it and
are left with it at the end of the computation, accounting for $2E_1
(x,y)$ irreversible bit operations, which is typically larger than
$E_3 (x,y)$. Up to additive logarithmic terms
 $E_1(x,y) \leq E_3 (x,y) \leq 2E_1 (x,y)$.
If the total computation time is limited
then the total number of irreversible bit operations will
rise. Resource-bounded versions of $E_3 (\cdot,\cdot )$
are studied in \cite{LiVi96}.

\paragraph{Thermodynamic Work.}
  Section \ref{s.thermo} considers the problem of defining a
thermodynamic entropy {\em cost} of transforming $x$ into $y$, and
argues that it ought to be an anti-symmetric, transitive function, in
contrast to the informational metrics which are symmetric.
  Landauer's principle connecting logical and physical
irreversibility is invoked to argue in favor of $K(x)-K(y)$ as the
appropriate (universal, anti-symmetric, and transitive) measure of the
thermodynamic work required to transform $x$ into $y$ by the most
efficient process.

\paragraph{Density in Information Metric Spaces.}
  Section \ref{s.dim} investigates the densities induced by the
optimal and sum information metrics.
  That is, how many objects are there within a given distance of a
given object.
  Such properties can also be viewed as ``dimensional'' properties.
They will govern many future applications of information distances.

\section{Kolmogorov Complexity}\label{s.props}

  Let $l(p)$ denote the length of the binary string $p$.
  Let $\# S$ denote the number of elements of set $S$.
  We give some definitions and basic properties of Kolmogorov
complexity.
  For all details and attributions we refer to
\cite{LiVitBook93}. There one can also find the
basic notions of computability theory and
Turing machines. The ``symmetry of information'' property
in Equation~\ref{e.addition} is from \cite{Gacs74}.
It refines an earlier version 
in \cite{ZvLe70} relating to the original
Kolmogorov complexity of \cite{Ko65}.

\begin{formal}{Definition}
  We say that a real-valued function $f(x,y)$ over strings or
natural numbers $x,y$ is {\em
upper-semicomputable} if the set of triples
 \[
  \setof{ (x,y,d):  f(x,y) < d,\mbox{ with } d \mbox{ rational}}
 \]
  is recursively enumerable.
  A function $f$ is {\em lower semicomputable} if $-f$ is 
upper-semicomputable.
\end{formal}

\begin{formal}{Definition}
  A {\em prefix set}, or prefix-free code, or prefix code, is a set of
strings such that no member is a prefix of any other member.
  A prefix set which is the domain of a partial recursive function
(set of halting programs for a Turing machine) is a special type of
prefix code called a {\em self-delimiting} code because there is an
effective procedure which reading left-to-right 
determines where a code word ends without
reading past the last symbol.
  A one-to-one function with a range that is a self-delimiting code
will also be called a self-delimiting code.
\end{formal}

We can map $\{0,1\}^*$ one-to-one onto the
natural numbers by associating each string with its index
in the length-increasing lexicographical ordering
\begin{equation}
( \epsilon , 0),  (0,1),  (1,2), (00,3), (01,4), (10,5), (11,6),
\ldots ,
\label{(2.1)}
\end{equation}
where $\epsilon$ denotes the empty word, that is, $l(\epsilon)=0$.
This way we have a binary representation for the
natural numbers that is different from the standard
binary representation.
It is convenient not to distinguish between the
first and second element of the same pair,
and call them ``string'' or ``number''
arbitrarily. As an example, we have $l(7)=00$.
A simple self-delimiting code we use throughout is obtained by reserving one
symbol, say 0, as a stop sign and encoding a
natural number $x$ as $1^x 0$.
We can prefix an object with its length and iterate
this idea to obtain ever shorter codes:
\begin{equation}
\label{ladder}
\lg_i (x)  = \left\{ \begin{array}{ll}
1^x 0 & \mbox{for $i=0$}, \\
\lg_{i-1} (l(x)) x & \mbox{for $i>0$}.
\end{array} \right.
\end{equation}
Thus, $\lg_1 (x) = 1^{l(x)} 0 x$ and has
length $l(\lg_1 (x)) = 2l(x) + 1$; 
$\lg_2 (x)  =  \lg_1 (l(x)) x$ and has length
$ l(\lg_2 (x) )  =  l(x) + 2l(l(x)) + 1$.
  From now on, we will denote by $\lea$ an inequality to within an
additive constant, and by $\eqa$ the situation when both $\lea$ and
$\gea$ hold.
  We will also use $\lel$ to denote an inequality to within an
additive logarithmic term, and $\eql$ to denote
the situation when both $\lel$ and
$\gel$ hold.
   Using this notation we have for example
 \[
  l(\lg_3 (x))\lea l(x)+\log l(x) + 2\log\log l(x).
 \]

  Define the pairing function
 \begin{equation}\label{e.ngl}
  \ngl{x,y}=\lg_2(x)y
 \end{equation}
  with inverses $\ngl{\cdot}_1,\ngl{\cdot}_2$.
  A partial recursive function $F(p,x)$ is called {\em
self-delimiting} if for each $x$,
 $\setof{p: F(p,x) < \infty }$ is a self-delimiting code.
(``$ F(p,x) < \infty$'' is shorthand for 
``there is a $y$ such that $F(p,x)=y$.'') 
  The argument $p$ is called a {\em self-delimiting program}
for $y:=F(p,x)$ from $x$, because, owing to the self-delimiting property, no
punctuation is required to tell the machine where $p$
ends and the input to the machine can be simply 
the concatenation $px$.

 \begin{formal}{Remark}
  Our results do not depend substantially on the use of
self-delimiting programs but for our purpose
 this form of the theory of Kolmogorov complexity is
cleaner and easier to use.
  For example, the simplicity of the normalization property in Section
\ref{s.axioms} depends on the self-delimiting property.
 \end{formal}
 
 \begin{formal}{Remark}\label{r.turing}
  Consider a multi-tape Turing machine $M$ with a distinguished
semi-infinite tape called the {\em program tape}.
  The program tape's head begins scanning the leftmost square of the
program.
  There is also an input tape and, possibly, a separate output tape
and work tapes.
  We say that $M$ computes the partial function $F(p,x)$ by a {\em
self-delimiting computation} if for all $p$ and $x$ for which
$F(p,x)$ is defined: 
 \begin{itemize}
\item $M$
with program $p$ and input $x$ halts with output
  $F(p,x)$
written on the output tape.
  \item The program tape head scans all of $p$
but not beyond $p$.
 \end{itemize}
  A partial recursive function is self-delimiting if and
only if there is a self-delimiting computation for it.
  A Turing machine performing a self-delimiting computation is
called a {\em self-delimiting Turing machine}.
 \end{formal}

  In what follows, informally, we will often call a self-delimiting
partial recursive function $F$ a {\em prefix machine} or {\em
self-delimiting machine} even though it is only the function computed
by such a machine.

\begin{formal}{Definition}\label{def.pkc}
  The {\em conditional descriptional complexity}, (the
``self-delimiting'' version) $K_F(y|x)$ of $y$ with condition $x$, with
respect to the machine $F$ is defined by
\[K_F(y|x) := \min \{l(p): F(p,x)=y  \}, \]
or $\infty$ if such $p$ do
not exist.
  There is a prefix machine $U$ (the universal self-delimiting 
Turing machine) with the
property that for every other prefix machine $F$ there
is an additive constant $c_F$ such that for all $x,y$
 \[ K_U(y|x) \le K_F(y|x) +
c_F. \]
  (A stronger property that is satisfied by many universal machines
  $U$ is that for all $F$ there is a string $s_F$ such that
  for all $x,y,p$ we have $U(s_F p, x)=F(p,x)$, from which the stated
  property follows immediately.)
  Since $c_F$ depends on $F$ but not on $x,y$
such a prefix machine $U$ will be called {\em optimal} or {\em universal}.
  We fix such an optimal machine $U$ as reference, write
 \[
  K(y|x) := K_U(y|x)
 \]
  and call $K(y|x)$ the conditional {\em Kolmogorov complexity} of $y$ with
respect to $x$. The unconditional Kolmogorov complexity of $y$ is defined
as $K(y) := K(y|\epsilon)$ where $\epsilon$ is the empty word.
\end{formal}

  We give a useful characterization of $K(y|x)$.
  It is easy to see that $K(y|x)$ is an upper-semicomputable function
with the property that for each $x$ we have
 \begin{equation}\label{e.Kraft}
 \sum_y 2^{-K(y|x)}\le 1.
 \end{equation}
Namely, for each $x$ the set of $K(y|x)$'s is a subset of
the length set of a prefix-code.
  Therefore property~\ref{e.Kraft} is a consequence of the so-called Kraft
inequality.
  It is an important fact that the function $K(y|x)$
is minimal with respect to the normalization
property~\ref{e.Kraft}: 

\begin{formalsl}{Lemma}\label{lem.usn}
For {\em every} 
upper-semicomputable function $f(x,y)$
satisfying $\sum_y 2^{-f(x,y)}\le 1$ we have $K(y|x)\lea f(x,y)$.
\end{formalsl}
A prominent example of such a function is the {\em algorithmic entropy}
\[
H(y|x) := - \log \sum_{p: U(p,x)=y} 2^{-l(p)} .
\]
Since $K(y|x)$ is the length of the shortest program $p$
such that $U(p,x)=y$ we have $K(y|x) \geq H(y|x)$, and
because $H(y|x)$ is upper-semicomputable and
satisfies $\sum_y 2^{-H(y|x)}\le 1$ (by the Kraft inequality)
 we have $K(y|x) \lea H(y|x)$.
Together this shows that $H(y|x) \eqa K(y|x)$ (almost all the entropy is
concentrated in the shortest program).

  The functions $\ngl{x,y,z}$, etc.\ are defined with the help of
$\ngl{x,y}$ in any of the usual ways.
  We introduce the notation
 \[
  K(x,y)=K(\ngl{x,y}),\ K(x|y,z)=K(x|\ngl{y,z}),\
 \]
  etc.
  Kolmogorov complexity has the following addition property:
 \begin{equation}\label{e.addition}
   K(x,y)\eqa K(x) + K(y|x,K(x)) .
 \end{equation}
  Ignoring for a moment the term $K(x)$ in the condition of the second
term of the right-hand side, this property says, analogously to the
corresponding property of information-theoretic entropy, that the
information content of the pair $(x,y)$ is equal to the information
content of $x$ plus the information needed to restore $y$ from $x$.

The {\em mutual information} between $x$ and $y$ is the
quantity
\begin{equation}\label{eq.mutualinfo}
   I(x:y) = K(x)+K(y)-K(x,y).
\end{equation}
This is the algorithmic counterpart of the mutual information between
two random variables $I(X:Y)=H(X)+H(Y)-H(X,Y)$.  Because of the
conditional $K(x)$
term in Equation~\ref{e.addition}, the usual relation between conditional
and mutual information holds only to within a logarithmic
error term
  (denoting $\ol x := \ngl{x,K(x)}$):
\begin{eqnarray*}
I(x:y) 
  & \eqa & K(x)-K(x|\ol y) 
 \eqa  K(y) - K(y| \ol x)  \\
 & = & K(x)-K(x|y)+O(\log(K(y)) =
K(y)-K(y|x)+O(\log(K(x)) .
\end{eqnarray*}
Thus, within logarithmic error, $I(x:y)$ represents both the
information in $y$ about $x$ and that in $x$ about $y$.  We consider $x$
and $y$ to be ``independent'' whenever  $I(x:y)$  is (nearly) zero.

  Mutual information should not be confused with ``common
information.''
  Informally, we can say that a string $z$ contains information common
in $x$ and $y$ if both $K(z|x)$ and $K(z|y)$ are small.
  If this notion is made precise it turns out that common information
is can be very low even if mutual information is
large~\cite{GacsKornComnInf73}.

\section{Max Distance}\label{s.conv}

  In line with the identification of the Kolmogorov complexity $K(x)$
as the information content of $x$, \cite{Ko65}, we define
 the information distance between $x$ and $y$ as the length
of the shortest program that converts $x$ to $y$ and $y$ to $x$.
The program itself is retained before, during, and after
the computation.
  This can be made formal as follows.
  For a partial recursive function $F$ computed by a prefix
(self-delimiting) Turing machine, let
 \[
  E_F (x,y) := \min \{l(p): F (p,x)=y, \: F (p,y)=x \}.
 \]
  There is a universal prefix machine $U$ (for example the reference machine 
in Definition~\ref{def.pkc}) such that
for every partial recursive prefix function $F$ and all $x,y$
  \[
  E_U (x,y) \leq E_F (x,y) + c_F ,
  \]
  where $c_F$ is a constant that depends on $F$ but not on
$x$ and $y$.
For each two universal prefix machines $U$ and
$U'$, we have for all $x,y$ that
 $|E_U (x,y)-E_{U'} (x,y)| \leq c$, with $c$ a constant
depending on $U$ and $U'$ but not on $x$ and $y$.
Therefore, with $U$ the reference universal prefix machine $U$ of 
Definition~\ref{def.pkc} we define
 \[
  E_0 (x,y) := \min \{l(p): U (p,x)=y, \: U (p,y)=x \}.
 \]
Then $E_0 ( \cdot , \cdot )$ is the universal effective information
distance which is clearly optimal and symmetric, and will be
shown to satisfy the triangle inequality.
  We are interested in the precise expression for $E_0$.

\subsection{Maximum overlap}

  The conditional complexity $K(y|x)$ itself is unsuitable as 
information distance because it is unsymmetric: $K(\epsilon|x)$, where
$\epsilon$ is the empty string, is small for all $x$, yet intuitively
a long random string $x$ is not close to the empty string.
  The asymmetry of the conditional complexity $K(x|y)$ can be remedied
by defining the informational distance between $x$ and $y$
to be the sum of the relative complexities, $K(y|x) + K(x|y)$.
  The resulting metric will overestimate the information required to
translate between $x$ and $y$ in case there is some redundancy between
the information required to get from $x$ to $y$ and the information
required to get from $y$ to $x$.

  This suggests investigating to what extent the information required
to compute $x$ from $y$ can be made to overlap with that required to
compute $y$ from $x$.
  In some simple cases, it is easy to see how complete overlap can be
achieved, so that the same minimal program suffices to compute $x$
from $y$ as to compute $y$ from $x$.
  A brief discussion of this and an outline of the results to follow
were given in Section~\ref{sect.intro}.

\begin{formal}{Definition}
  The {\em max distance} $E_1$ between $x$ and $y$ is defined
by
  \[
  E_1 (x,y) \\ := \max \{K(x|y),K(y|x) \}.
  \]
\end{formal}

By definition of Kolmogorov complexity, every program $p$ that
computes $y$ from $x$ and also computes $x$ from $y$ satisfies
$l(p) \geq E_1(x,y)$, that is, 
\begin{equation}\label{eq.e0e1}
E_0(x,y) \geq E_1 (x,y).
\end{equation}
In Theorem~\ref{t.excess} we show that this relation also holds the other way:
$E_0 (x,y) \leq E_1 (x,y)$ up to an additive logarithmic term.
Moreover, the information to compute from
$x$ to $y$ can always be maximally correlated with
the information to compute from $y$ to $x$.
  It is therefore never the case that a large amount of information is
required to get from $x$ to $y$ and a large {\em but independent}
amount of information is required to get from $y$ to $x$.

\begin{formalsl}{Conversion Theorem}\label{t.diff}\label{t.excess}
Let $K(x|y) = k_1$ and $K(y|x)=k_2$, and $l = k_2 - k_1 \geq 0$.
There is a string $d$ of length $l$ and a string $q$ of length 
\[k_1 + K(k_1,k_2) + O(1) \]
 such that $U(q,xd)=y$ and $U(q,y)=xd$. 
\end{formalsl}
\begin{proof}
Given $k_1,k_2$, we can enumerate the set
$S=\{(x,y): K(x|y) \leq k_1, K(y|x) \leq k_2 \}$.
Without loss of generality, assume that $S$ is enumerated without repetition,
and with witnesses of length exactly $k_1$ and $k_2$.
Now consider a dynamic graph $G=(V,E)$
where $V$ is the set of binary strings,
and $E$ is a dynamically growing set of edges that starts out empty.
 
Whenever a pair $(x,y)$ is enumerated, we add an edge
$e=\{xd, y\}$ to $E$. Here, $d$ is chosen to be the $(i2^{-k_1})$th binary
string of length $l$,
where $i$ is the number of times we have enumerated a pair with $x$ as the
first element. So the first $2^{k_1}$ times we enumerate a pair $(x, \cdot)$ we
choose $d=0^l$, for the next $2^{k_1}$ times we choose $d=0^{l-1}1$, etc.
The condition $K(y|x) \leq k_2$ implies that $i<2^{k_2}$ hence
$i2^{-k_1} < 2^l$, so this choice is well-defined.
 
In addition, we ``color'' edge $e$ with a binary string of length $k_1 + 3$.
 Call two edges {\em adjacent} if they have a common endpoint.
If $c$ is the minimum color not yet appearing on any edge adjacent to either
$xd,x,yd$ or $y$, then $e$ is colored $c$. Since the degree of every node
is bounded by $2^{k_1}$ (when acting as an $xd$) plus $2^{k_1}$ (when acting
as a $y$), a color is always available.
(This particular color assignment is needed in the proof of 
Theorem~\ref{theo.e0ee1}.)
 
  A {\em matching}\index{matching} is a maximal set of nonadjacent edges.
Note that the colors partition $E$ into at most $2^{k_1 + 3}$ matchings,
since no edges of the same color are ever adjacent. Since the pair $(x,y)$
in the statement of the theorem is necessarily enumerated, there is some
$d$ of length $l$ and color $c$ such that the edge $\{xd,y\}$ is added to $E$
with color $c$.
 
Knowing $k_1,k_2,c$ and either of the nodes $xd$ or
$y$, one can dynamically reconstruct $G$, find the unique $c$-colored edge
adjacent to this node, and output the neighbour. Therefore,
a self-delimiting program $q$ of size 
$K(k_1,k_2) + k_1 + O(1)$ suffices to compute
in either direction between $xd$ and $y$.
\end{proof}

The theorem states that $K(y|xd,q), K(xd|y,q) \eqa 0$.
  It may be called the {\em Conversion Theorem} since it
asserts the existence of a difference string $q$ that converts both
ways between $xd$ and $y$ and at least one of these conversions is
optimal.
  If $k_1=k_2$, then $d= \epsilon$ and
the conversion is optimal in both directions.

\begin{formalsl}{Theorem}\label{theo.e0ee1}
  Assume the notation above. Then, with $\eql$
denoting equality up to additive logarithmic terms:
\begin{eqnarray*}
E_0(xd,y) & \eql & E_1(xd,y) \;(\eql l(q)) \\
E_0 (x,y) &  \eql & E_1 (x,y) 
\; (\eql l(qd)) .
\end{eqnarray*}
\end{formalsl}

\begin{proof}
(First displayed equation)
Assume the notation and proof of Theorem~\ref{t.diff}.
First note that $l(q) \eql E_1 (xd,y)$.
Moreover, $q$ computes between $xd$ and
$y$ in both directions 
and therefore $l(q) \geq E_0(xd,y)$ by the 
minimality of $E_0 (\cdot , \cdot)$. Hence
$E_1 (xd,y) \gel E_0 (xd,y)$. Together with Equation~\ref{eq.e0e1}
this shows 
the first displayed equation holds.

(Second displayed equation) This requires
an extra argument to show that
the program $p := qd$ is a program to compute between $x$
and $y$ in both directions. Namely,
knowing $k_1,k_2,c,d$ and string $x$
one can dynamically reconstruct $G$  and find the first
enumerated $c$-colored edge
adjacent to either node $x$ or node $xd$ and output the neighbour
($yd$ or $y$ respectively). 
By a similar argument as in the previous case 
we now
obtain the second displayed equation.
\end{proof}

 \begin{formal}{Remark}
  The same proofs work for the non-self-delimiting
Kolmogorov complexity as in \cite{Ko65} and would
also give rise to a logarithmic correction term in the theorem.
 \end{formal}

\begin{formal}{Remark}\label{r:orth}
  The difference program $p = qd$ 
in the above theorem is independent of
$x$ in the sense that the mutual information
$I(p:x)$ as defined in Equation~\ref{eq.mutualinfo} is
nearly 0.
  This follows from $K(x)+K(p)=K(x,y)+O(\log K(x))$ (use
  Equation~\ref{e.addition} with $K(y|x)=K(p)$).
  The program $p$ is at the same time completely dependent on the pair $(x,y)$.

  If $k_1=k_2$ then $d = \epsilon$ and $p =q$. Then $p=q$ is a conversion
program from $x$ to $y$ and from $y$ to $x$ and it is both independent
of $x$ and independent of $y$, that is, $I(p:x), I(p:y)$ are
both nearly $0$.
  The program $p$ is at the same time completely dependent on the pair $(x,y)$.
 \end{formal}

\begin{formal}{Remark (Mutual Information Formulation)}\label{r:mutinfo}
 Let us reformulate the result of this section in terms
of mutual information as defined in Equation~\ref{eq.mutualinfo}.
Let $p$ be a shortest program transforming $x$ to $y$  and
let $q$ be a shortest program transforming $y$ to $x$.
We have shown that $p$ and $q$ can depend on each other
as much as possible: the mutual information in $p$ and $q$ is
maximal:
$I(p:q)= \min \{l(p),l(q)\}$ up to an additive $O( \log I(p:q) )$ term.
\end{formal}

  \subsection{Minimum overlap}\label{sect.minoverlap}
This section can be skipped at first reading; the material
is difficult and it is not used
in the remainder of the paper.
  For a pair $x,y$ of strings, we found that shortest program
$p$ converting $x$ into $y$ and $q$ converting $y$ into $x$ 
can be made to overlap maximally. 
In Remark~\ref{r:mutinfo} this result is formulated in
terms of mutual information.
The opposite question is
whether $p$ and $q$ can always be made completely {\em independent}, that is,
can we choose $p$ and $q$ such that $I(p:q)=0$?
hat is,
is it true that for every $x,y$ there are $p,q$ such that
 $K(p)=K(y|x)$, $K(q)=K(x|y)$, $I(p:q)=0$, $U(p,x)=y$, $U(q,y)=x$,
where the first three equalities hold up to an
additive $O( \log I(p:q) )$ term.
This is evidently true in case 
$x$ and $y$ are random with respect to one another, that is, $K(x|y) \geq l(x)$
and $K(y|x) \geq l(y)$. 
Namely, without loss of generality let $y=uv$ with $l(u)=l(x)$.
We can choose $p:= (x \oplus u)v$ as a shortest program that
computes from $x$ to $y$ and $q := x \oplus u$ as
a  shortest program that computes from $y$ to $x$, and therefore obtain
maximum overlap $I(p:q)= \min \{l(p),l(q)\}$.
However, we can also choose shortest programs 
 $p := y$ and $q := x$ to realize minimum overlap
$I(p:q)=0$. 
The question arises whether we can {\em always} 
choose $p,q$ with $I(p:q)=0$ even when $x$ and $y$ are not random
with respect to one another.

\begin{formal}{Remark}
N.K. Vereshchagin suggested replacing 
``$I(p:q)=0$'' (that is, $K(p,q)=K(p)+K(q)$)
by ``$K(q|x)=0, K(p|y)=0$,''
everything up to an additive $O( \log I(p:q) )$ term. 
Then an affirmative answer to the
latter question would imply an affirmative answer to the former
question. 
\end{formal}

Here we study a related but formally different question: replace
the condition ``$I(p:q)=0$'' by ``$p$ is a function of only $y$'' 
and ``$q$ is a function of only $x$.'' Note that when this
new condition is satisfied 
it can still
happen that $I(p:q) > 0$. We may choose to ignore the latter
type of mutual information.

 We show that for every pair of integers $k_1,k_2 \geq 0$
there exists a function $f$ with $K(f) = k_1+k_2 + O(\log(k_1+k_2))$
such that
for every $x,y$ such that $K(x) \leq k_1, K(y|x) \leq k_2$ 
we have $K(y|x,f(y),f)=O(\log(k_1+k_2))$
and $l(f(y)) \approx k_2$,
that is, $f(y)$ has about $k_2$ bits and suffices together
with a description of $f$ itself to restore $y$ from
every $x$ from which this is possible using this many bits.
Moreover, there is no significantly simpler function $f$,
say $K(f|y) \ll \min \{ k_1,k_2\}$, with this property.

Let us amplify the meaning of this for
the question of the conversion programs
having low mutual information. First we need some terminology.
  When we say that $f$ is a {\em simple function} of $y$ we mean that
$K(f|y)$ is small.

  Suppose we have a minimal program $p$, of length $k_2$, converting $x$
to $y$ and a minimal program $q$ of length $k_1$ converting $y$ to
$x$.
  It is easy to see, just as in Remark \ref{r:orth} above that $y$ is
independent of $q$.
  Also, any simple function of $y$ is independent of $q$.
  So, if $p$ is a simple function of $y$, then it is independent of $q$.
  The question whether $p$ can be made a simple function of $y$ is
interesting in itself since it would be a generalization of the
Slepian-Wolf Theorem (see~\cite{CsiszarKorner80}).
  And it sounds no less counterintuitive at first than that theorem.
  If it were true then for {\em each} $y$ 
there is a $k_2$-bit program $p$
such that for {\em every} $x$ satisfying $K(y|x)\le k_2$, we can
reconstruct $y$ from the pair $(x,p)$.
  As stated already, we will show that
$p$ can be made a {\em
function} of $y$ independent of $x$;
  but we will also show that $p$
{\em cannot} be made a {\em simple} function of $y$.

  Before proceeding with the formal statement and proof
 we introduce a combinatorial
lemma.
  In a context where a partition $V=\bigcup_j V_j$ of a set $V$ is called
a coloring we say that two elements have {\em the same color} if they
belong to the same set $V_j$.

 \begin{formalsl}{Coloring Lemma}\label{lem.zf}
  On a set $V$, let us be given a set system with $M$ sets $S_i$
(possibly overlapping) of
size at most $N$ each.
  For $B>0$, a {\em $B$-coloring} of this system is a partition
$V=\bigcup_j V_j$ such that $\#(S_i\bigcap V_j)\le B$ for every $i,j$, 
that is, there are at
most $B$ points of the same color in a set $S_i$.
  There is a $B$-coloring with not more colors than
 \[
   (N/B) e(MN)^{1/B}.
 \]
 \end{formalsl}

\begin{formal}{Remark}
  Notice that $N/B$ colors are trivially required (and suffice
if the $S_i$'s are pairwise disjoint).
\end{formal}

 \begin{proof}
  If $B=N$ then one color is enough, so assume $B<N$.
  Let us try to color with $nN/B$ colors and then see what choice of
$n$ satisfies our needs.
  We choose the color of each element of $V$ independently, 
with a uniform distribution among the given number of colors, with
probability $p :=B/(nN)$.
  For each $i,j$, we can upperbound the probability that $\#(S_i\bigcap
V_j)>B$, using the Chernoff bound (see e.g.~\cite{CsiszarKorner80})
for large deviations in the law of large numbers.
  In application to the present case, this bound says that if in an
experiment of $N$ coin-tosses 
the success probability is $p$  then
for every $p'>p$, the probability that there are more than $Np'$
successes is at most $e^{cN}$ with
 \[
   c = p'\ln\frac{p}{p'} + (1-p')\ln\frac{1-p}{1-p'}.
 \]
(The $N$ coin tosses correspond to random coloring the elements in $S_i$ 
where a success is the coloring of an element with a given color
like ``blue.'')

  We apply this bound with $p=B/(nN)$ and $p'=B/N$. Summing over 
all sets (there are $M$ sets) and all colors used in each set (there are
at most $N$ colors used to color a set) we obtain that
  $MNe^{cN}$ upperbounds the probability that the random coloring
is not a $B$-coloring.
  Let us see what choice of $n$ makes this bound less than 1.

  Estimating the second term of the right-hand side above by
 $\ln x\le x-1$, it is at most $p'-p<p'$, hence
 $c<p'(\ln(p/p')+1)=(B/N)(-\ln n+1)$.
  Now the condition $MNe^{cN}<1$ turns into $\ln(MN) + Nc<0$.
  Substituting the above estimate for $c$, we get a stronger condition
$\ln(MN)+B \le B\ln n$, satisfied by $\ln n=(\ln(MN))/B+1$.
 \end{proof}

 \begin{formalsl}{Theorem}\label{t.Slepian-Wolf}
  {\rm (i)} There is a recursive function $R$
such that for every pair of integers $k_1,k_2 > 0$ there is
an integer $m$ with $\log m \leq k_1 + k_2$ and an integer
$b$ with $b \lea \log (k_1 + k_2)+ 2 \log \log (k_1 + k_2)$ 
such that for all $x,y$
with $K(x) \leq k_1$ and $K(y|x)\leq k_2$ 
    \[  K(y| x,f(y),m) \le b , \]
 where $f(y) := R(k_1,k_2,m,y)$ 
with $l(f(y)) \lea  k_2 $.

  {\rm (ii)} Using the notation in (i), even
allowing for much larger $b$ we cannot significantly eliminate
the conditional information $m$ required in (i):
If $b$ satisfies
 \begin{equation}\label{e.b-bound}
  0 \leq b < k_1 - 5 \log (k_1 + k_2),
 \end{equation}
then  every $m$
satisfying the conditions in {\rm (i)} also satisfies
 \[
  l(m) \geq k_2 -b  - 5 \log (k_1 + k_2).
 \]
 \end{formalsl}

 \begin{formal}{Remark}
  Thus, 
the extra information in $y$ needed in addition to $x$ to
restore $y$ can be made a function $f(y)$ of just $y$, and its
minimality implies that it will be essentially independent of $x$.
  However, there is a catch: it is indispensible for these results
that certain fixed oracle string $m$ describing how to
compute $f$ is also used in the
transformations.
  The role of this oracle string is to make the complexity function
computable over the set of strings of interest.
 \end{formal}

 \begin{formal}{Remark}
If also $K(y) \leq k_2$ then the theorem holds symmetrically
in $x$ and $y$.
  This is the sense in which the shortest programs $f(y)$ and $f(x)$,
converting $x$ into $y$ and $y$ into $x$, can be made
``non-overlapping'': they will be independent of the strings
they convert {\em from}.
 \end{formal}

 \begin{proof}
(i)  We first show the existence of $R$ and $m$ with the above
properties.
 As in the proof of Theorem \ref{t.diff},
let $G=(V,E)$ be a graph with the node set $V \subseteq \{0,1\}^*$
and $E$ consisting of those edges
$(x,y)$ with $K(x) \leq k_1$ and $K(y|x)\leq k_2$.
Let
 \begin{eqnarray*}
     M   &=& 2^{k_1},\ N = 2^{k_2};
 \\  S_x &=& \setof{y:  (x,y)\in E};
 \\  B   &=& k_1 + k_2;
 \\  m   &=& \#E.
 \end{eqnarray*}
  Then $\# S_x \leq N$, and the number of $x's$ with nonempty $S_x$
is at most $M$.
  According to the Coloring Lemma~\ref{lem.zf}, 
there is a $B$-coloring of the $M$ sets $S_x$ with at
most
 \begin{equation}\label{eq.numcs}
   (N/B)e(MN)^{1/B}=2eN/B
 \end{equation}
  colors.
  Let $R$ be a recursive function computing
a color $f(y)=R(k_1,k_2,m,y)$. 
  Using the numbers $k_1,k_2,m$ it reconstructs the graph $G$.
  Then it finds (if there is no better way, by exhaustive search) a
$B$-coloring of the $S_x$'s set system.
  Finally, it outputs the color of $y$.
  
  Let us estimate $K(y|x,f(y),m)$. 
Without loss of generality
we can assume that the
representation of $m \leq 2^{k_1 + k_2}$ is padded up to length exactly
$k_1 + k_2$. The logarithm of the number of colors is 
$\lea k_2 - \log (k_1 + k_2)$ so with padding we can represent
color $f(y)$ by a string of precisely that length. Therefore,
we can retrieve $k_1,k_2$ from the representations of $m$ and $f(y)$
in the conditional.
  Now for every $y\in S_x$, if we are given $k_1,k_2,m,x,$ and $f(y)$ then
we can list the set of
all $y$'s in $S_x$ with color $f(y)$. 
  Since the size of this list is at most $B$, the program to determine
$y$ in it needs only the number of $y$ in the enumeration, with a
self-delimiting code of length 
$l(\lg_2 (B))\lea \log(k_1+k_2) + 2 \log \log(k_1+k_2)$ with
$\lg_2$  as in Definition~\ref{ladder}.

(ii)  
Suppose that there is a number $m$ with the desired
properties with representation length 
 \begin{equation}\label{e.n-bound}
  l(m) < k_2 -b - 5 \log (k_1 + k_2),
 \end{equation}
 and $b$ satisfies \ref{e.b-bound}.
  We will arrive from here at a contradiction.
  First note that the number of $y$'s satisfying 
$K(y|x) \leq k_2$ for some $x$ with $K(x) \leq k_1$ 
as required in the theorem is
 \begin{equation}\label{eq.numys}
  \log\# \bigcup_x S_x \gea k_1 + k_2 - 2.2 \log(k_1+k_2).
 \end{equation}
  Namely, concatenating an arbitrary binary string $x$ with $K(x) \lea k_1$
and an arbitrary string $v$  with $K(v) \lea k_2$ we can form $y=xv$
and we have $K(y|x) \lea K(v) \lea k_2$.
This includes every $x$ with $l(x) \lea k_1 - 1.1 \log k_1$ and
every $v$ with $l(v) \lea
k_2 - 1.1 \log k_2$.
For appropriate additive constants in $\lea$ 
it will be true that for every such $x$, all such
strings $y$ will belong to $S_x$.

Choose an arbitrary recursive function $R$ satisfying
the statements of the theorem and Equation~\ref{e.n-bound}.
  For each possible value $c$ of $f(y)$
(where $f(y) := R(k_1,k_2,m,y)$), let 
\[ Y_c :=\setof{y: f(y)=c}. \]
Because the number of $y$'s is lower-bounded by Equation~\ref{eq.numys}
and  the size of $f(y)$ is upper-bounded by $l(f(y)) \lea k_{2}$
there is a $c$ such that
\begin{equation}\label{eq.lnc}
  \log\# Y_c \gea k_1 -  2.2 \log (k_1 + k_2 ).
\end{equation}
  Let $l$ be the first such $c$ found when enumerating all the sets
$Y_c$. This enumeration can be done as follows: Using $k_1$ we
enumerate all $x$ with $K(x) \leq k_1$ by running all programs of
length $\leq k_1$ in rounds of one step per program; when a program
halts its output is the next $x$ enumerated. For all of the enumerated $x$'s,
we use $k_2$ to enumerate all $y$'s with $K(y|x) \leq k_2$ in a similar fashion.
Finally, for each enumerated $y$ compute $f(y) = R(k_1,k_2,m,y)$ and
enumerate the $Y_c$'s.

   Therefore, given the recursive function $R$, the integers
$k_1,k_2,m$, and an constant-length program we can
enumerate the $Y_c$'s,
determine $l$, and 
enumerate $Y_l$.
We can describe $R$ by a constant-length self-delimiting program and
the integers $k_1,k_2,m$ by
a self-delimiting program $\mu := \lg_3 (k_1) \lg_3 (k_2) \lg_3 (m)$
with $\lg_3$ as in Definition~\ref{ladder}.
  Then, for every $i$ such that $y_i$ is the $i$-th element in this enumeration
of $Y_l$:
 \begin{eqnarray*}
  K(y_i) &\lea& l( \mu ) + \log i + 1.1\log\log i
\\         &\lea & l(m ) + \log i + 4.4\log(k_1 +k_2).
 \end{eqnarray*}
  If
\begin{equation}\label{eq.iy}
\log i < k_2 - l(m) - 4.5\log(k_1 +k_2)
\end{equation} 
and $k_1 +k_2$ is large
enough, then for every $x$ we have
 \[
  K(y_i |x) \leq K(y_i) + O(1) \le k_2 .
 \]
  Let $t=\min \{ k_1 ,k_2 -l(m)\}$.
By Equations~\ref{eq.lnc}, \ref{eq.iy}, for every $x$ 
  there are at least 
\[ 2^{t-4.5\log(k_1 +k_2)}\]
 values of
$i$ with $K(y_i |x )\leq k_2$.
  Then, for every $x$ there must be
at least one of these $y_i$'s, say $y$, that satisfies
 \[ K(y|x,f(y),m )\ge t-4.5\log(k_1 +k_2). \]
This follows trivially by counting the number of programs
of length less than $t-4.5\log(k_1 +k_2)$.
  Hence, by the property $b \geq K(y|x,f(y),m )$ 
assumed in the statement of the
theorem: 
 \[
  b \geq \min \{k_1 ,k_2-l(m)\} -4.5\log(k_1 +k_2).
 \]
  If $k_1 < k_2 - l(m)$ then this contradicts \ref{e.b-bound}, otherwise it
contradicts \ref{e.n-bound}.
 \end{proof}

\section{Cognitive Distance}\label{s.axioms}

  Let us identify digitized black-and-white pictures with binary strings.
  There are many distances defined for binary strings.
  For example, the Hamming distance and the Euclidean distance.
  Such distances are sometimes appropriate.
  For instance, if we take a binary picture, and change a few bits on
that picture, then the changed and unchanged pictures have small
Hamming or Euclidean distance, and they do look similar.
  However, this is not always the case.
  The positive and negative prints of a photo have the largest
possible Hamming and Euclidean distance, yet they look similar to us.
  Also, if we shift a picture one bit to the right, again the Hamming
distance may increase by a lot, but the two pictures remain similar.
  Many approaches to pattern recognition try to define pattern
similarities with respect to pictures, language sentences,
vocal utterances, and so on. Here we assume that similarities 
between objects can
 be represented by effectively computable functions
(or even upper-semicomputable functions)
of binary strings. This seems like a minimal prerequisite
for machine pattern recognition
and physical cognitive processes in general.
  Let us show that the distance $E_1$ defined above is, in a sense,
minimal among all such reasonable similarity measures.

  For a cognitive similarity metric the metric requirements 
  do not suffice: a distance measure like $D(x,y) = 1$
for all $x \neq y$ must be excluded.
  For each $x$ and $d$, we want only finitely many elements $y$ at a
distance $d$ from $x$.
  Exactly how fast we want the distances of the strings $y$ from $x$
to go to $\infty$ is not important: it is only a matter of scaling.
  In analogy with Hamming distance in the space of binary sequences,
it seems natural to require that there should not be more than $2^d$
strings $y$ at a distance $d$ from $x$.
  This would be a different requirement for each $d$.
  With prefix complexity, it turns out to be more convenient to
replace this double series of requirements (a different one for each
$x$ and $d$) with a single requirement for each $x$:
 \[
  \sum_{y: y \neq x} 2^{-D(x,y)}<1 .
 \]
  We call this the {\em normalization property} since a certain sum is
required to be bounded by 1.

  We consider only distances that are computable in some broad sense.
  This condition will not be seen as unduly restrictive.
  As a matter of fact, only upper-semicomputability of $D(x,y)$ will
be required.
  This is reasonable: as we have more and more time to process $x$ and
$y$ we may discover more and more similarities among them, and thus
may revise our upper bound on their distance.
  The upper-semicomputability means exactly that $D(x,y)$ is the limit
of a computable sequence of such upper bounds.

\begin{formal}{Definition}
  An {\em admissible distance} $D(x,y)$ is a total nonnegative
function on the pairs $x,y$ of binary strings that is 0 if and only if
$x=y$, is symmetric, satisfies the triangle inequality, is 
upper-semicomputable and normalized, that is, it is an upper-semicomputable,
normalized, metric.  An admissible distance $D(x,y)$
is {\em universal} if for every admissible distance $D'(x,y)$ we have
$D(x,y) \lea D'(x,y)$.
\end{formal}
  The following theorem shows that $E_1$ is a universal
(that is, optimal) admissible distance.
  We find it remarkable that this distance happens to also have a
``physical'' interpretation as the approximate length of the
conversion program of Theorem \ref{t.diff}, and, as shown in the
next section, of the smallest program that transforms $x$ into
$y$ on a reversible machine.

\begin{formalsl}{Theorem}\label{t.optimal.cognitive.dist}
  For an appropriate constant $c$, let $E(x,y)=E_1(x,y)+c$ if
$x\not=y$ and 0 otherwise.
  Then $E(x,y)$ is a universal admissible metric.
  That is, it is an admissible distance
and it is minimal in the sense that for every admissible
distance $D(x,y)$ we have
 \[
  E(x,y)\lea D(x,y).
 \]
\end{formalsl}

\begin{proof}
  The nonnegativity and symmetry properties are immediate from the
definition.
  To prove the triangle inequality, let $x,y,z$ be given and
assume, without loss of generality, that $E_1(x,z)=K(z|x)$.
  Then, by the self-delimiting property (or, the easy direction of the
addition property),

 \begin{eqnarray*}
   E_1(x,z) &=&    K(z|x)\lea K(y,z|x)\lea K(y|x)+K(z|x,y)
\\          &\lea& K(y|x)+K(z|y) \le E_1(x,y) + E_1(y,z).
 \end{eqnarray*}
  Hence there is a nonnegative integer constant $c$ such that
 $E_1(x,z) \le E_1(x,y) + E_1(y,z) +c$.
  Let this $c$ be the one used in the statement of the theorem, then
$E(x,y)$ satisfies the triangle inequality without an additive
constant.

  For the normalization property, we have
  \[
  \sum_{y: y \neq x} 2^{-E_1(x,y)}\le \sum_{y: y \neq x} 2^{-K(y|x)}\le 1 .
 \]
  The first inequality follows from the definition of $E_1$, and the
second one follows from \ref{e.Kraft}.

  The minimality property follows from the characterization of
$K(y|x)$ given after \ref{e.Kraft}.
  This property says that if $f(x,y)$ is an upper-semicomputable
function with $\sum_{y: y \neq x} 2^{-f(x,y)}\le 1$ then $K(y|x)\lea f(x,y)$.
  This implies that for every admissible distance
$D(\cdot , \cdot )$ we have both $K(y|x)\lea D(x,y)$ and $K(x|y)\lea D(y,x)$.
 \end{proof}
\begin{formal}{Remark (Universal Cognitive Distance)}
  The universal admissible distance $E_1$ minorizes {\em all}
admissible distances: if two pictures are $d$-close under some
admissible distance, then they are $ \lea d$-close under this universal
admissible distance.
  That is, the latter discovers all effective feature similarities
or cognitive similarities
between two objects: it is the universal cognitive similarity metric.
\end{formal}

\section{Reversible Computation Distance}\label{s.rev}

  Reversible models of computation in which the transition function
is one-to-one have been explored especially in connection with the
question of the thermodynamic limits of computation.
  Reversible Turing machines were introduced by
Lecerf~\cite{Lecerf63}, and independently but much later by
Bennett~\cite{Bennett73,Bennett82}.
  Further results concerning them can be found in
\cite{Bennett82,Bennett89,LevineSherman90,LiVi96}.

Consider the standard model of Turing machine.
The elementary operations are rules
in quadruple format $(p,a,b,q)$ meaning that a machine in
state $p$ scanning symbol $a$ writes a symbol or moves the 
scanning head one square left, one square right, or not at all
(as indicated by $b$) and enters state $q$. 

Quadruples are said
to {\em overlap in domain} if they cause the machine in the same state
and scanning the same symbol to perform different actions.
A {\em deterministic Turing machine} is defined as a Turing machine
with quadruples that pairwise do not overlap in domain.
 
Now consider a special format (deterministic) Turing
machines using quadruples of two
types: {\em read/write} quadruples and {\em move} quadruples.
A read/write quadruple $(p,a,b,q)$ causes the machine in state
$p$ scanning tape symbol $a$ to write symbol $b$ and enter state $q$.
A move quadruple $(p,\perp,\sigma ,q)$ causes the machine
in state $p$ to move its tape head by $\sigma \in \{-1,0,+1\}$
squares and enter state $q$,
oblivious to the particular symbol in the currently scanned tape square.
(Here ``$-1$'' means ``one square left,'' ``$0$'' means ``no move''
and ``$+1$'' means ``one square right.'') Quadruples are said
to {\em overlap in range} if they cause the machine to enter
the same state and either both write the same symbol or
(at least) one of them moves the head. Said differently,
quadruples that enter the same state overlap in range
unless they write different symbols.
A {\em reversible Turing machine} is a deterministic Turing machine
with quadruples that pairwise do not overlap in range.
A $k$-tape reversible Turing machine uses $(2k+2)$ tuples
that for each tape separately, select a read/write or move on that
tape. 
Moreover, every pair of tuples having the same initial state must 
specify differing scanned symbols on at least one tape (to guarantee
non-overlapping domains), and every pair of tuples having the same
final state must write differing symbols on at least one tape (to
guarantee non-overlapping ranges).
 
To show that each partial recursive function can be computed
by a reversible Turing machine one can proceed as follows.
Take the standard irreversible Turing machine computing that function.
We modify it by
adding an auxiliary storage tape called the ``history tape.''
The quadruple rules are extended to 6-tuples to additionally
manipulate the history tape.
To be able to reversibly undo (retrace)
the computation deterministically, the new 6-tuple
rules have the effect that the machine keeps a record
on the auxiliary history tape consisting of
the sequence of quadruples executed on the original tape.
Reversibly undoing a computation
entails also erasing the record
of its execution from the history tape.
 
This notion of reversible computation means
that only one-to-one recursive functions can be computed.
To reversibly simulate $t$ steps of an
irreversible computation from $x$ to $f(x)$
one reversibly  computes from input $x$
to output $\langle x, f(x) \rangle$.
Say this takes $t' = O(t)$ time.
Since this reversible simulation at some time instant
has to record the entire
history of the irreversible computation, its space use increases
linearly with the number of simulated steps $t$. That is,
if the simulated irreversible computation uses $s$ space, then
for some constant $c > 1$ the simulation uses
$t'\approx c+ct$ time and $s'\approx c + c(s+t)$ space.
After computing from $x$ to $f(x)$
the machine reversibly copies $f(x)$, reversibly undoes
the computation from  $x$ to $f(x)$ erasing its history tape
in the process, and ends with one copy of $x$ and
one copy of $f(x)$ in the format $\langle x, f(x) \rangle$
and otherwise empty tapes.

  Let $\psi_i$ be the partial recursive function computed by the
$i$'th such {\em reversible Turing machine}.
  We let $\phi_i$ denote the partial recursive function
computed by the $i$'th ordinary (in general irreversible) Turing
machine.
  Among the more important properties of reversible Turing machines
are the following \cite{Bennett82,Bennett89,LevineSherman90}:
 \begin{description}
  \item[Universal reversible machine]
  There is a universal reversible machine, i.e.~an index $u$ such that
for all $k$ and $x$, $\psi_u(\ngl{k,x})=\ngl{k,\psi_k(x)}$.
  \item[Irreversible to reversible]
  Two irreversible algorithms, one for computing $y$ from $x$ and the
other for computing $x$ from $y$, can be efficiently combined to
obtain a reversible algorithm for computing $y$ from $x$.
  More formally, for any two indices $i$ and $j$ one can effectively
obtain an index $k$ such that, for any strings $x$ and $y$, if
$\phi_i(x)=y$ and $\phi_j(y)=x$, then $\psi_k(x) =y$.
  \item[Saving input copy]
From any index $i$ one may obtain an index $k$ such that
$\psi_k$ has the same domain as $\phi_i$ and, for every
$x$,  $\psi_k(x) = \ngl{x,\phi_i(x)}$.
  In other words, an arbitrary Turing machine can be simulated by a
reversible one which saves a copy of the irreversible machine's input
in order to assure a global one-to-one mapping.
  \item[Efficiency]
  The above simulation can be performed rather efficiently.
  In particular, for any $\epsilon>0$ one can find a reversible
simulating machine which runs in time $O(T^{1+\epsilon})$ and space
$O(S\log (T/S))$ compared to the time $T$ and space $S$ of the
irreversible machine being simulated.
  \item[One-to-one functions]
  From any index $i$ one may effectively obtain an index $k$ such that
if $\phi_i$ is one-to-one, then $\psi_k=\phi_i$.
  The reversible Turing machines $\{\psi_k\}$, therefore, provide a
G\"{o}del-numbering of all one-to-one partial recursive functions.
 \end{description}
  The connection with thermodynamics comes from the fact that in
principle the only thermodynamically costly computer operations are
those that are {\em logically irreversible}, i.e.~operations that map
several distinct logical states of the computer onto a common
successor, thereby throwing away information about the computer's
previous state
 \cite{Landauer61,Bennett73,FredkinToffoli82,Bennett82,LiVi96}.
  The thermodynamics of computation is discussed further in Section
\ref{s.thermo}.
  Here we show that the minimal program size for a reversible
computer to transform input $x$ into output $y$ is equal within an
additive constant to the size of the minimal conversion string $p$ of
Theorem \ref{t.diff}.

  The theory of reversible minimal program size is conveniently
developed using a reversible analog of the universal self-delimiting
function (prefix machine) $U$ defined in Section \ref{s.props}.
\begin{formal}{Definition}
  A partial recursive function $F(p,x)$ is called a {\em reversible
self-delimiting function} if
 \begin{description}
 \item for each $p$, $F(p,x)$ is one-to-one as a function of $x$;
 \item for each $x$, $\setof{p: \exists y\, F(p,x) = y }$ is a
prefix set;
 \item for each $y$, $\setof{p: \exists x\, F(p,x)=y}$ is a
prefix set.
\end{description}
\end{formal}
 \begin{formal}{Remark}
  A referee asked whether the last two of these conditions can be
replaced with the single stronger one saying that $\setof{p: \exists
x,y\, F(p,x)=y}$ is a prefix set.
  This does not seem to be the case.
 \end{formal}

  In analogy with Remark \ref{r.turing}, we can define the notion of a
{\em reversible self-delimiting computation} on a reversible Turing
machine.
  Take a reversible multi-tape Turing machine $M$ with a special
semi-infinite read-only tape called the {\em program tape}.
  There is now no separate input and output tape, only an input-output
tape.
  At the beginning of the computation, the head of the program tape is
on the starting square.

  We say that $M$ computes the partial function $F(p,x)$ by a {\em
reversible self-delimiting computation} if for all $p$ and $x$
for which $F(p,x)$ is defined:
 \begin{itemize}
  \item 
  $M$ halts with output $y:=F(p,x)$ written on its output tape
 performing a one-to-one mapping $x \leftrightarrow y$ on the
input-output tape under the control of the program $p$.
  \item The program tape head scans all of $p$
  but never scans beyond the end of $p$.
  \item At the end of the computation, the program tape head rests on
the starting square.
   Once it starts moving backward it never moves forward again.
  \item Any other work tapes used during the computation are supplied
in blank condition at the beginning of the computation and must be
left blank at the end of the computation.
 \end{itemize}

  It can be shown (see the references given above) that a function $F$
is reversible self-delimiting if and only if it can be computed by a
reversible self-delimiting computation.
  Informally, again, we will call a reversible self-delimiting
function also a {\em reversible self-delimiting (prefix) machine}.

  A {\em universal reversible prefix machine} $\UR$, which is optimal
in the same sense of Section \ref{s.props}, can be shown to exist, and
the {\em reversible Kolmogorov complexity} $\KR(y|x)$ is defined as
\[\KR(y|x) :=  \min\setof{l(p):\UR(p,x)=y}. \]

  In Section \ref{s.conv}, it was shown that for any strings $x$
and $y$ there exists a conversion program $p$, of length at most
logarithmically greater than
 \[
  E_1 (x,y) = \max\{K(y|x),K(x|y)\}
 \]
  such that $U(p,x)=y$ and $U(p,y)=x$.
  Here we show that the length of this minimal such conversion program
is equal within a constant to the length of the minimal {\em
reversible} program for transforming $x$ into $y$.

\begin{formalsl}{Theorem}\label{t.rev.pg.len}
 \[
  \KR(y|x) \eqa \min\setof{l(p): U(p,x)=y,\ U(p,y)=x}.
 \]
\end{formalsl}

\begin{proof}
  ($\gea$) The minimal reversible program for $y$ from $x$, with
constant modification, serves as a program for $y$ from $x$ for the
ordinary irreversible prefix machine $U$, because reversible
prefix machines are a subset of ordinary prefix machines.
We can reverse
a reversible program by adding an $O(1)$ bit prefix program
to it saying ``reverse the following program.''

  ($\lea$) The proof of the other direction is an example of the general
technique for combining two irreversible programs, for $y$ from $x$
and for $x$ from $y$, into a single reversible program for $y$ from
$x$.
  In this case the two irreversible programs are the same,
since by Theorem \ref{t.diff} the minimal conversion program $p$ is
both a program for $y$ given $x$ and a program for $x$ given $y$.
  The computation proceeds by several stages as shown in 
Figure~\ref{fig.rev.concat}.
  To illustrate motions of the head on the self-delimiting program
tape, the program $p$ is represented by the string ``prog'' in the
table, with the head position indicated by a caret.

  Each of the stages can be accomplished without using any
many-to-one operations.
\begin{figure}[t]
\begin{center}
{\small
  \begin{tabular}{l|c r c c c }
  {\sc Stage and Action}
  & {\sc Program} & \multicolumn{4}{c}{{\sc Work Tape}}
\\ \hline \hline
\\ 0. Initial configuration
  & \^{p}rog     & $x$  &                 &    &
\\ 1. Compute $y$, saving history
  & pro\^{g}     & $y$  & $(y|x)$-history &     &
\\ 2. Copy $y$ to blank region
  & pro\^{g}     & $y$  & $(y|x)$-history & $y$ &
\\ 3. Undo comp.\ of $y$ from $x$
  & \^{p}rog     & $x$ &                 & $y$ &
\\ 4.  Swap $x$ and $y$
  & \^{p}rog     & $y$  &                 & $x$ &
\\ 5.  Compute $x$, saving history
  & pro\^{g}     & $x$  & $(x|y)$-history & $x$ &
\\ 6.  Cancel extra $x$
  & pro\^{g}     & $x$  & $(x|y)$-history &     &
\\ 7.  Undo comp.  of $x$ from $y$
  & \^{p}rog     & $y$ &                 &     &
\end{tabular}
}
\caption{Combining irreversible computations of $y$ from $x$ and $x$ from
$y$ to achieve a reversible computation of $y$ from $x$}
\label{fig.rev.concat}
\end{center}
\end{figure}
 
In stage 1, the computation of $y$ from $x$, which might otherwise
involve irreversible steps, is rendered reversible by saving a
history, on previously blank tape, of all the information that would
have been thrown away.
 
In stage 2, making an extra copy of the output onto blank tape is
an intrinsically reversible process, and therefore can be done
without writing anything further in the history.
Stage 3 exactly undoes the work of stage 1, which is possible
because of the history generated in stage 1.
 
Perhaps the most critical stage is stage 5, in which $x$ is
computed from $y$ for the sole purpose of generating a history of
that computation.
  Then, after the extra copy of $x$ is reversibly disposed of in stage
6 by cancelation (the inverse of copying onto blank tape), stage 7
undoes stage 5, thereby disposing of the history and the remaining
copy of $x$, while producing only the desired output $y$.
 
  Not only are all its operations reversible, but the computations
from $x$ to $y$ in stage 1 and from $y$ to $x$ in stage 5 take place
in such a manner as to satisfy the requirements for a reversible
prefix interpreter.
  Hence, the minimal irreversible conversion program $p$, with
constant modification, can be used as a reversible program for $\UR$
to compute $y$ from $x$.
  This establishes the theorem.
\end{proof}

\begin{formal}{Definition}
  The {\em reversible distance} $E_2(x,y)$ between $x$ and $y$ 
is defined by
\[
 E_2(x,y) := \KR(y|x) = \min\setof{l(p) : \UR(p,x)=y}.
 \]
\end{formal}
  As just proved, this is within an additive constant of the size of
the minimal conversion program of Theorem \ref{t.diff}.
  Although it may be logarithmically greater than the optimal
distance $E_1$, it has the intuitive advantage of being the actual
length of a concrete program for passing in either direction between
$x$ and $y$.
  The optimal distance $E_1$ on the other hand is defined only as the
greater of two one-way program sizes, and we don't know whether
it corresponds to the
length of any two-way translation program.

  $E_2 (x,y)$ may indeed be legitimately called a distance because it is
symmetric and obeys the triangle inequality to within an additive
constant (which can be removed by the additive rescaling technique
used in the proof of Theorem \ref{t.optimal.cognitive.dist}).

\begin{formalsl}{Theorem}
 \[
  E_2(x,z)  \lea  E_2(x,y) + E_2(y,z)
 \]
\end{formalsl}

\begin{proof}
  We will show that, given reversible $\UR$ programs $p$ and $q$, for
computing $(y|x)$ and $(z|y)$ respectively, a program of the form
$spq$, where $s$ is a constant supervisory routine, serves to compute
$z$ from $x$ reversibly.
  Because the programs are self-delimiting, no punctuation is needed
between them.
  If this were an ordinary irreversible $U$ computation, the
concatenated program $spq$ could be executed in an entirely
straightforward manner, first using $p$ to go from $x$ to $y$,
then using $q$ to go from $y$ to $z$.
  However, with reversible $\UR$ programs, after executing $p$,
the head will be located at the beginning of the program tape,
and so will not be ready to begin reading $q$.
  It is therefore necessary to remember the length of the first
program segment $p$ temporarily, to enable the program head to
space forward to the beginning of $q$, but then cancel this
information reversibly when it is no longer needed.

  A scheme for doing this is shown in Figure~\ref{fig.rev.cancel}, 
where the program
tape's head position is indicated by a caret.
  To emphasize that the programs $p$ and $q$ are strings concatenated
without any punctuation between them, they are represented
respectively in the table by the expressions ``pprog'' and ``qprog'',
and their concatenation $pq$ by ``pprogqprog''.
  
  Notice that transcribing ``pprog'' in stage 1 is straightforward:
as long as the program tape head moves forward such a transcription
will be done; according to our definition of reversible
self-delimiting computation above, this way the whole program will be
transcribed.

\begin{figure}\centering
\begin{tabular}{ l|c c c}
 Stage and Action                          & Program tape &
 \multicolumn{2}{c}{Work Tape}
\\ \hline
\\ 0. Initial configuration                & \^{p}progqprog & $x$ &
\\ 1. Compute $(y|x)$, transcribing pprog. & \^{p}progqprog & $y$ & pprog
\\ 2. Space forward to start of qprog.     & pprog\^{q}prog & $y$ & pprog
\\ 3. Compute $(z|y)$.                     & pprog\^{q}prog & $z$ & pprog
\\ 4. Cancel extra pprog as head returns.  & \^{p}progqprog & $z$ &
\end{tabular}
\caption{Reversible execution of concatenated programs for $(y|x)$ and $(z|y)$
to transform $x$ into $z$.}
\label{fig.rev.cancel}
\end{figure}

\end{proof}

\section{Sum Distance}\label{s.alt-dist}

Only the irreversible
erasures of a computation need to dissipate energy.
This raises the question
of the minimal amount of irreversibility
required in transforming string $x$ into string $y$,
that is, the number of bits we have to add to $x$ at the beginning
of a reversible computation from $x$ to $y$,
and the number of garbage bits
left (apart from $y$) at the end of the computation that
must be irreversibly erased to obtain a ``clean'' $y$.

  The reversible distance $E_2$ defined in the previous section,
is equal to the length of a ``catalytic'' program, which allows
the interconversion of $x$ and $y$ while remaining unchanged
itself.
  Here we consider noncatalytic reversible computations which
consume some information $p$ besides $x$, and produce some
information $q$ besides $y$.

  Even though consuming and producing information may seem to be
operations of opposite sign, we can define a distance $E_3 (\cdot , \cdot)$
 based on the
notion of information flow, as the minimal {\em sum} of amounts of extra
information flowing into and out of the computer in the course of
the computation transforming $x$ into $y$.
  This quantity measures the number of irreversible bit operations in
an otherwise reversible computation.
The resulting distance turns out to be within
a logarithmic additive term of the sum of the conditional
complexities $ K(y|x)+K(x|y)$. See \cite{LiVi96}
for a more direct proof than the one provided here, and for a study 
of resource-limited (for example with respect to time) measures
of the number of irreversible bit operations. For our treatment here
it is crucial that computations can take unlimited time and space
and therefore $E_3 ( \cdot, \cdot )$ represents a limiting quantity
that cannot be realized by feasible computation.
  For a function $F$ computed by a reversible Turing machine, define
 \[
  E_F (x,y) :=
 \min\setof { l(p)+ l(q):  F (\ngl{p,x})=\ngl{q,y } }.
 \]
 \begin{formal}{Remark}
  Since $p$ will be consumed it would be too awkward and not
worth the trouble to try to extend the notion of self-delimiting for
this case; so, the computations we consider will not be
self-delimiting over $p$.
 \end{formal}
 
  It follows from the existence of universal reversible Turing
machines mentioned in Section \ref{s.rev} that there is a universal
reversible Turing machine $\UR'$ (not necessarily
self-delimiting) such that for all functions $F$
computed on a reversible Turing machine, we have
 \[
  E_{\UR'} (x,y) \leq E_{F} (x,y) + c_{F}
 \]
 for all $x$ and $y$, where $c_F$ is a constant which depends on
$F$ but not on $x$ or $y$.

\begin{formal}{Remark}
  In our definitions we have pushed all bits to be irreversibly
provided to the start of the computation and all bits to be 
irreversibly erased to
the end of the computation.
  It is easy to see that this is no restriction.
  If we have a computation where irreversible acts happen throughout
the computation, then we can always mark the bits to be 
irreversibly erased,
waiting with actual erasure until the end of the computation.
  Similarly, the bits to be provided can be provided (marked) at the
start of the computation while the actual reading of them
(simultaneously unmarking them) takes place throughout the
computation.

  By Landauer's principle, which we meet in Section~\ref{s.thermo},
the number of irreversible bit erasures in a computation gives a lower
bound on the unavoidable energy dissipation of the computation, each
bit counted as $kT \ln 2$, where $k$ is Boltzmann's constant and $T$
the absolute temperature in degrees Kelvin.
  It is easy to see (proof of
Theorem~\ref{theorem.sumd}) that the minimal number of garbage bits
left after a reversible computation going from $x$ to $y$ 
is about $K(x|y)$ and in
the computation from $y$ to $x$ it is about $K(y|x)$.
 \end{formal}

\begin{formal}{Definition}
  We fix a universal reference reversible Turing machine $\UR'$.
The {\em sum distance} $E_3 (x,y)$ is defined by 
 \[
   E_3(x,y): = E_{\UR'}(x,y).
 \]
\end{formal}

\begin{formalsl}{Theorem}\label{theorem.sumd}
 \[
  E_3(x,y)=K(x|y)+K(y|x)+O(\log (K(x|y)+K(y|x))).
 \]
\end{formalsl}

\begin{proof}
  ($\geq$) We first show the lower bound $E_3(x,y) \geq K(y|x)+K(x|y)$.
  Let us use the universal prefix machine $U$ of Section
\ref{s.props}.
  Due to its universality, there is a constant-length
binary string $r$ such that for
all $p,x$ we have
 \[
   U(r\lg_2(p),x)=\ngl{\UR'(\ngl{p,x})}_2
 \]
  (The function $\lg_2$ in Definition~\ref{ladder} makes $p$ self-delimiting.
  Recall that $\ngl{ \cdot, \cdot}_2$ selects the second element of the pair.)
  Suppose $\UR'(\ngl{p,x})=\ngl{q,y}$.
  Then it follows that $y=U(r\lg_2(p),x)$, hence
 \[
   K(y|x)\lea l(r \lg_2(p)) \lea  l(\lg_2(p)) \lea l(p)+2\log l(p).
 \]
  Since the computation is reversible, the garbage information
$q$ at the end of the computation yielding $\ngl{y,q}$ serves the 
r\^ole of program when we reverse the computation to compute
$x$ from $y$.
  Therefore, we similarly have $K(x|y) \lea l(q)+2\log l(q)$, which
finishes the proof of the lower bound.

 ($\leq$) Let us turn to the upper bound and assume $k_1=K(x|y) \le
k_2=K(y|x)$ with $l=k_2-k_1 \geq 0$.
  According to Theorem \ref{t.excess}, there is a string $d$ of length
$l$ such that
 $K(xd|y) \eqa k_1+ K(k_1,k_2)$ and $K(y|xd) \eqa k_1+ K(k_1,k_2)$. 
  According to Theorem \ref{t.diff} and Theorem \ref{t.rev.pg.len}
there is a self-delimiting program $q$ of 
length $\eqa k_1+ K(k_1,k_2)$ going reversibly
between $xd$ and $y$.
  Therefore with a constant extra program $s$, the universal
reversible machine will go from  $qxd$ to $qy$.
  And by the above estimates
 \[
  l(qd)+l(q)\lea 2k_1+l + 2 K(k_1,k_2) = k_1+k_2+ O(\log k_2).
 \]
 \end{proof}

  Note that all bits supplied in the beginning to the computation, apart
from input $x$, as well as all bits erased at the end of the
computation, are {\em random} bits.
  This is because we supply and delete only shortest programs, and a
shortest program $q$ satisfies $K(q) \geq l(q)$, that is, it is
maximally random.

\begin{formal}{Remark}
It is easy to see that up to an additive logarithmic 
term the function $E_3 (x,y)$ is a metric on $\{0,1\}^*$;
in fact it is an admissible (cognitive) distance as defined in 
Section~\ref{s.axioms}.
\end{formal}
 
\section{Relations Between Information Distances}

   The metrics we have considered can be arranged in increasing
order.
  As before, the relation $\lel$ means inequality to within an additive
$O(\log)$, and $\eql$ means $\lel$ and $\gel$.
\begin{eqnarray*}
          && E_1(x,y) =\max\{K(y|x),K(x|y)\}
 \\       && \eql E_2(x,y) = \KR(y|x)
 \\       && \eqa E_0(x,y) = \min\setof{l(p):U(p,x)=y,\ U(p,y)=x}
 \\       && \lel K(x|y)+K(y|x) \eql E_3(x,y)
 \\       && \lel 2 E_1(x,y).
\end{eqnarray*}
  The sum distance $E_3$, is tightly bounded between the optimum
distance $E_1$ and twice the optimal distance.
   The lower bound is achieved if one of the conditional complexities
$K(y|x)$ and $K(x|y)$ is zero, the upper bound is reached if the two
conditional complexities are equal.

  It is natural to ask whether the equality $E_1(x,y)\eql E_2(x,y)$
can be tightened.
  We have not tried to produce a counterexample but the answer is
probably no.

\section{Thermodynamic Cost}\label{s.thermo}

  Thermodynamics, among other things, deals with the amounts of heat
and work ideally required, by the most efficient process, to convert
one form of matter to another.
  For example, at 0~C and atmospheric pressure, it takes 80 calories
of heat and no work to convert a gram of ice into water at the same
temperature and pressure.
  From an atomic point of view, the conversion of ice to water at 0~C
is a reversible process, in which each melting water molecule gains
about 3.8 bits of entropy (representing the approximately
$2^{3.8}$-fold increased freedom of motion it has in the liquid state),
while the environment loses 3.8 bits.
  During this ideal melting process, the entropy of the universe
remains constant, because the entropy gain by the ice is compensated
by an equal entropy loss by the environment.
  Perfect compensation takes place only in the limit of slow melting,
with an infinitesimal temperature difference between the ice and the
water.

Rapid melting, e.g.~when ice is dropped into hot water, is
thermodynamically irreversible and inefficient, with the the hot water
losing less entropy than the ice gains, resulting in a net and
irredeemable entropy increase for the combined system. (Strictly speaking, the
microscopic entropy of the universe as a whole does not increase,
being a constant of motion in both classical and quantum mechanics. 
Rather what happens when ice is dropped into hot water is that the
marginal entropy of the (ice + hot water) system increases, while the
entropy of the universe remains constant, due to a growth of mutual
information mediated by subtle correlations between the (ice + hot
water) system and the rest of the universe.  In principle these
correlations could be harnessed and redirected so as to cause the warm
water to refreeze, but in practice the melting is irreversible.)

   Turning again to ideal reversible processes, the entropy change in
going from state $X$ to state $Y$ is an anti-symmetric function of $X$
and $Y$; thus, when water freezes at 0~C by the most efficient
process, it gives up 3.8 bits of entropy per molecule to the
environment.
  When more than two states are involved, the entropy changes are
transitive: thus the entropy change per molecule of going from ice to
water vapor at 0~C ($+32.6$ bits) plus that for going from vapor to
liquid water ($-28.8$ bits) sum to the entropy change for going from ice
to water directly.
  Because of this asymmetry and transitivity, entropy can be
regarded as a thermodynamic potential or state function: each state
has an entropy, and the entropy change in going from state $X$ to
state $Y$ by the most efficient process is simply the entropy
difference between states $X$ and $Y$.

  Thermodynamic ideas were first successfully applied to computation
by Landauer.
  According to {\em Landauer's principle}
 \cite{Landauer61,Bennett82,Zurek89a,Zurek89b,Caves90}
 an operation that
maps an unknown state randomly chosen from among
$n$ equiprobable states onto a known common successor state
must be accompanied by an
entropy increase of $\log_2 n$ bits in other, non-information-bearing
degrees of freedom in the computer or its environment.
  At room temperature, this is equivalent to the production of
 $kT \ln 2$ (about $7\cdot 10^{-22}$) calories of waste heat per bit
of information discarded. 

The point here is the change from
``ignorance'' to ``knowledge'' about the state, that is, the gaining
of information and not the erasure in itself (instead
of erasure one could consider measurement that would make
the state known).

  Landauer's priniciple follows from the fact that such a logically
irreversible operation would otherwise be able to decrease the
thermodynamic entropy of the computer's data without a compensating
entropy increase elsewhere in the universe, thereby violating the
second law of thermodynamics.

  Converse to Landauer's principle is the fact that when a computer
takes a physical {\em randomizing} step, such as tossing a coin, in
which a single logical state passes stochastically into one of $n$
equiprobable successors, that step can, if properly harnessed, be
used to remove $\log_2 n$ bits of entropy from the computer's
environment.
  Models have been constructed, obeying the usual conventions of
classical, quantum, and thermodynamic thought-experiments
 \cite{Landauer61,KeyesLandauer70,Bennett73,Bennett82}
 \cite{FredkinToffoli82,Landauer82,Likharev82,Benioff82,Feynman85}
 showing both the ability in principle to perform logically
reversible computations in a thermodynamically reversible fashion
(i.e.~with arbitrarily little entropy production), and the ability
to harness entropy increases due to data randomization within a
computer to reduce correspondingly the entropy of its
environment.

  In view of the above considerations, it seems reasonable to assign
each string $x$ an effective thermodynamic entropy equal to its
Kolmogorov complexity $K(x)$.
  A computation that erases an $n$-bit random string would
then reduce its entropy by $n$ bits, requiring an entropy increase in
the environment of at least $n$ bits, in agreement with Landauer's
principle.

  Conversely, a randomizing computation that starts with a string
of $n$ zeros and produces $n$ random bits has, as its typical
result, an algorithmically random $n$-bit string $x$, i.e.~one for
which $K(x) \approx n$.
   By the converse of Landauer's principle, this randomizing
computation is capable of removing up to $n$ bits of entropy from the
environment, again in agreement with the identification of the
thermodynamic entropy and Kolmogorov complexity.

  What about computations that start with one 
(randomly generated or unknown) string
$x$ and end with another string $y$?
  By the transitivity of entropy changes one is led to say that the
thermodynamic cost, i.e.~the minimal entropy increase in the environment,
of a transformation of $x$ into $y$, should be
 \[
  W(y|x) = K(x)-K(y),
 \]
  because the transformation of $x$ into $y$ could be thought of as a
two-step process in which one first erases $x$, then allows $y$ to be
produced by randomization.
  This cost is obviously anti-symmetric and transitive, but is not even
semicomputable.
 Because it involves the {\em difference} of two
semicomputable quantities, it is at best expressible as the {\em non-monotone}
limit of a computable sequence of approximations.
  Invoking the identity~\cite{Gacs74} $K(x,y) \eqa K(x) +
K(y|x^*)$, where $x^*$ denotes the first minimal program for $x$
in enumeration order (or equivalently, $x^* := \ngl{x,K(x)}$), the
above cost measure $W(y|x)$ can also be interpreted as a
difference in conditional complexities,
 \[
  W(y|x) \eqa K(x|y^*)-K(y|x^*) \:.
 \]
  Such indirect conditional complexities, in which the input string is
supplied as a minimal program rather than directly, have been
advocated by Chaitin~\cite{Chaitin75} on grounds of their similarity to
conditional entropy in standard information theory.

  An analogous anti-symmetric cost measure based on the difference of
direct conditional complexities
 \[
  W'(y|x) =K(x|y)-K(y|x).
 \]
  was introduced and compared with $W(x|y)$ by Zurek~\cite{Zurek89a},
who noted that the two costs are equal within a logarithmic additive
term. 
  Here we note that $W'(y|x)$ is non-transitive to a similar extent.

  Clearly, $W'(y|x)$ is tied to the study of distance $E_3$, the sum of
irreversible information flow in and out of the computation.
  Namely, analysis of the proof of Theorem~\ref{theorem.sumd} shows
that up to logarithmic additional terms, a necessary and sufficient
number of bits of $K(y|x)$ (the program) needs to be supplied at the
start of the computation from $x$ to $y$, while a necessary and
sufficient number of bits of $K(x|y)$ (the garbage) needs to be
irreversibly erased at the end of the computation.
  The thermodynamical analysis of Landauer's principle at the
beginning of this section says the thermodynamic cost, and hence
the attending heat dissipation, of a computation of $y$ from $x$
is given by the number of irreversibly erased bits minus
the number of irreversibly provided bits, that is, $W'(y|x)$.



  It is known that there exist strings~\cite{Gacs74} $x$ of each
length such that $K(x^*|x)\approx \log l( x)$, where $x^*$ is the
minimal program for $x$.
  According to the $W'$ measure, erasing such an $x$ via the
intermediate $x^*$ would generate $\log l(x)$ less entropy than
erasing it directly, while for the $W$ measure the two costs would be
equal within an additive constant.
  Indeed, erasing in two steps would cost only
$K(x|x^*)-K(x^*|x)+K(x^*|0)-K(0|x^*)\eqa K(x)-K(x^*|x)$ while erasing
in one step would cost $K(x|0)-K(0|x)=K(x)$.

  Subtle differences like the one between $W$ and $W'$ pointed out
above (and resulting in a slight nontransitivity of $W'$) depend on
detailed assumptions which must be, ultimately, motivated by
physics~\cite{Zurek89b}.
  For instance, if one were to follow Chaitin~\cite{Chaitin75} and
define a ${\it Kc}$-complexity as ${\it Kc}(x):=K(x), {\it Kc}(x,y):=K(x,y)$ but the 
conditional information ${\it Kc}(y|x) := K(y|x^*)$
then the joint information would be given directly by ${\it Kc}(x,y) \eqa {\it
Kc}(x) +
{\it Kc}(y|x)$, and the ${\it Kc}$-analogues ${\it Wc}'(y|x) = {\it Wc}(y|x)$ 
would hold without logarithmic
corrections (because ${\it Kc}(y|x)={\it Kc}(y|x^*)$).
  This ${\it Kc}$ notation is worth considering especially because the
joint and conditional {\it Kc}-complexities satisfy
equalities which also obtain for the statistical entropy
(i.e.~Gibbs-Shannon entropy defined in terms of probabilities) without
logarithmic corrections.
  This makes it a closer analog of the thermodynamic entropy.
  Moreover---as discussed by Zurek~\cite{Zurek89b}, in a cyclic
process of a hypothetical Maxwell demon-operated engine involving
acquisition of information through measurement, expansion, and
subsequent erasures of the records compressed by reversible
computation---the optimal efficiency of the cycle could be assured
only by assuming that the relevant minimal programs are already
available.

  These remarks lead one to consider a more general issue of
entropy changes in nonideal computations.
  Bennett~\cite{Bennett82} and especially Zurek~\cite{Zurek89b} have
considered the thermodynamics of an intelligent demon or engine which
has some capacity to analyze and transform data $x$ before erasing it.
  If the demon erases a random-looking string, such as the digits of
$\pi$, without taking the trouble to understand it, it will commit a
thermodynamically irreversible act, in which the entropy of the data
is decreased very little, while the entropy of the environment
increases by a full $n$ bits.
  On the other hand, if the demon recognizes the redundancy in $\pi$,
it can transform $\pi$ to an (almost) empty string by a reversible
computation, and thereby accomplish the erasure at very little
thermodynamic cost. See for a comprehensive treatment \cite{LiVitBook93}.

  More generally, given unlimited time, a demon could approximate the
semicomputable function $K(x)$ and so compress a string $x$ to size
$K(x)$ before erasing it.
  But in limited time, the demon will not be able to compress $x$ so
much, and will have to generate more entropy to get rid of it.
  This tradeoff between speed and thermodynamic efficiency is
superficially similar to the tradeoff between speed and efficiency
for physical processes such as melting, but the functional form of
the tradeoff is very different.
  For typical physical state changes such as melting, the excess
entropy produced per molecule goes to zero inversely in the time $t$
allowed for melting to occur.
  But the time-bounded Kolmogorov complexity $K^t(x)$, i.e.~the size
of the smallest program to compute $x$ in time less than $t$, in
general approaches $K(x)$ only with uncomputable slowness as a
function of $t$ and $x$. These issues have been analyzed in more detail
by two of us in \cite{LiVi96}.

\section{Density Properties}\label{s.dim}

  In a discrete space with some distance function, the rate of growth
of the number of elements in balls of size $d$ can be considered as a
kind of ``density'' or ``dimension'' of the space.
  For all information distances one significant feature is how many
objects there are within a distance $d$ of a given object.
  From the pattern recognition viewpoint such information tells how
many pictures there are within the universal admissible (max)
distance $E_1(x,y)=d$.
  For the reversible distance $E_2(x,y) =d$ this tells us how many
objects one can reach using a reversible program of length $d$.
  For the sum distance $E_3(x,y) =d$ this tells us how many objects
there are within $d$ irreversible bit operations of a given object.

  Recall the distances $E_1(x,y)=\max\{K(x|y),K(y|x)\}$ and
$E_3(x,y)\eql K(x|y)+K(y|x)$.
  For a binary string $x$ of length $n$, a nonnegative number $d$ 
  and $i=1,3$, let
$B_i(d,x)$ be the set of strings $y \neq x$ with $E_i(x,y)\le d$, and
$B_i(d,x,n) :=B_i(d,x)\bigcap \{0,1\}^n$.
 
  The functions $B_i(d,x)$ behave rather simply:
$\log\# B_i(d,x)$ grows essentially like $d$.
  The functions $B_i(d,x,n)$ behave, however, differently.
  While $\log\# B_1(d,x,n)$ grows essentially like $d$, the function
$\log\# B_3(d,x,n)$ grows essentially like $d/2$.
  This follows from the somewhat more precise result in
\ref{t.thermo.random} below.
First we treat the general case below that says that
balls around $x$ of radius $d$ with $d$ random with respect to $x$
contain less elements:
{\em neighborhoods of tough radius's contain less neighbors}.

 \begin{formalsl}{Theorem}\label{t.dim}
  Let $x$ be a binary string of length $n$. The number of binary
strings $y$ with $E_1 (x,y) \leq d$ satisfies
 \begin{eqnarray*}
   &&  \log  \# B_1(d,x) \eqa d - K(d|x) ; \\
   && d-K(d) \lea \log  \# B_1(d,x,n) \lea d - K(d|x) .
 \end{eqnarray*}
  The last equation holds only for $n \geq d-K(d)$:
  for $n < d-K(d)$ we have $ \log \# B_1(d,x,n) \eqa n$.
 \end{formalsl}

\begin{proof}
($B_1(d,x) \lea $) For every binary string $x$ 
\begin{eqnarray*}
\sum_{d=0}^{\infty} \# B_1(d,x) 2^{-d-1} & = &
\sum_{d=0}^{\infty} \sum_{j=0}^d 2^{-d+j-1} \sum_{y:E_1 (x,y)=j \& y \neq x} 
2^{-j} \\
& = &
\sum_{d=0}^{\infty} \sum_{j=0}^d 2^{-d+j-1} \sum_{y: E_1(x,y)=j \&  y \neq
x} 2^{-E_1 (x,y)} \\
& = &
\sum_{i=1}^{\infty} 2^{-i} \sum_{y: y \neq x} 2^{-E_1 (x,y)} 
 \leq 1,
\end{eqnarray*}
where the last inequality follows from the properties of $E_1 (\cdot , \cdot )$
proven in Theorem~\ref{t.optimal.cognitive.dist}.
Since $f(x,d) := \log (  2^{d+1} / \# B_1(d,x))$ is upper-semicomputable
and satisfies $\sum_{d} 2^{-f(x,d)} \leq 1$, by Lemma~\ref{lem.usn}
we have $K(d|x) \lea f(x,d) \eqa d - \log \# B_1 (d,x)$.

($B_1 (d,x) \gea $) For all $i<2^{d-K(d|x)}$, 
  consider the strings $y_i= \lg_3 (i) x$ where $\lg_3$ 
  is the self-delimiting code of Definition~\ref{ladder}.
  The number of such strings $y_i$ is $2^{d-K(d|x)}$.
  Clearly, for every $i$, we have $K(x|y_i) \eqa 0$
  and $K(y_i|x) \eqa K(i|x)$. Therefore,
\[ E_1(x,y_i) \lea  K(i|x) . \]
Each $i$ can be represented by a string $z_i$ of length precisely
$d-K(d|x)$, if necessary by padding it up to this length.
Let $q$ be a shortest self-delimiting program computing $d$ from $x$.
By definition $l(q)=K(d|x)$. The program $qz_i$ is a self-delimiting
program to compute $i$ from $x$: Use $q$ to compute $d$ from $x$ and
subsequently use $d-l(q)=d-K(d|x)= l(z_i)$ to determine where $z_i$ ends.
Hence, $K(i|x) \lea l(qz_i)=d$ from which $E_1 (x,y_i) \lea d$ follows. 
The implied additive constants in $\lea$ 
can be removed in any of the usual ways.

($B_1(d,x,n) \lea$) 
Since $\# B_1 (d,x,n) \leq \# B_1(d,x)$ the upper bound on the latter
is also an upper bound on the former.

($B_1 (d,x,n) \gea $ and $n \geq d-K(d)$) 
For the $d-K(d)$ lower bound on $\log \# B_1(d,x,n)$ the proof
is similar but now we consider all $i < 2^{d-K(d)}$ and we choose 
the strings $y_i=x \oplus i$ where $\oplus$ means 
bitwise exclusive-or (if $l(i) < n$ then assume that the missing 
bits are $0$'s).

($B_1 (d,x,n) $ and $n < d-K(d)$) in that case we obtain all
strings in $\{0,1\}^n$ as $y_i$'s in the previous proof.
 \end{proof}

Note that $K(d) \lea \log d + 2 \log \log d$.
  It is interesting that a similar dimension relation holds also for
the larger distance $E_3(x,y) \eql K(y|x)+K(x|y)$.

 \begin{formalsl}{Theorem}\label{theo.balle3}
  Let $x$ be a binary string.
The number $ \# B_3(d,x)$ of binary
strings $y$ with $E_3 (x,y) \leq d$ satisfies
 \[
 \log\# B_3(d,x) \eql d-K(d|x).
 \]
 \end{formalsl}

 \begin{proof}
  ($\lea$) 
  This follows from the previous theorem since $E_3\ge
E_1$.

( $\gel$)
  Consider strings $y$ of the form $px$ where $p$
is a self-delimiting program.
  For all such programs, $K(x|y)\eqa 0$, since $x$ can be
recovered from $y$ by a constant-length program.
  Therefore $E_3(x,y)\eql K(y|x) \eqa K(p|x)$.
  Now just as in the argument of the previous proof, there are at least
$2^{d-K(d|x)}$ such 
strings $p$ with $K(p|x)\le d$.
 \end{proof}

  The number of strings of length $n$  within any $E_3$-distance of
a {\em random} string $x$ of length $n$, 
(that is, a string with $K(x)$ near $n$)
turns out to be different from the number of strings
of length $n$ within the same $E_1$-distance.
  In the $E_3$-distance: ``{\em tough guys have few
  neighbors of their own size}''. 

  In particular, a random string $x$ of length $n$ has only about
$2^{d/2}$ strings of length $n$ within $E_3$-distance $d$ while
there are essentially $2^{d}$ such strings within
$E_1$-distance $d$ of $x$ by Theorem~\ref{t.dim}.
  Moreover, since Theorem~\ref{theo.balle3}
  showed that every string has essentially $2^d$ neighbors
  altogether in $E_3$-distance $d$, for every random string $x$
  asymptotically {\em almost all} its neighbors within $E_3$-distance
  $d$ have {\em length unequal} $n$.
  The following theorem describes the general situation.

\begin{formalsl}{Theorem}\label{t.thermo.random}
  For each $x$ of length $n$ we have 
 \[
  \log\# B_3(d,x,n) \eql \frac{n+d -K(x)}{2},
 \]
  while $n-K(x) \leq d$.
  (For $n-K(x) > d$ we have $\log\# B_3(d,x,n) \eql d $.)
 \end{formalsl}

 \begin{proof}
  Let $K(x) \eql n-  \delta (n)$ (for example, $K(x) \eqa n +K(n) - \delta
  (n)$.

\noindent($\geq$)
  Let $y=x^*z$ with $l(y)=n$ and $l(z)=\delta (n)$,
 and let $x^*$ be the first
self-delimiting
program for $x$ ($l(x^*)=K(x)$) that we 
find by dovetailing all computations on
programs of length less than $n$.
  We can retrieve $z$ from $y$ using at most $O(\log n)$ bits.
  There are $2^{\delta (n)}$ different such $y$'s.
  For each such $y$ we have $K(x|y)=O(1)$, since $x$ can be
retrieved from $y$ using $x^*$.
  Now suppose that we also replace the fixed first $l/2$ bits
of $y$ by an arbitrary $u \in \{0,1\}^{l/2}$ for some
value of $l$ to be determined later. 
  Then, the total number of $y$'s increases to
 $2^{\delta (n) + l/2}$.

  These choices of $y$ must satisfy $E_3(x,y) \leq d$.
  Clearly, $K(y|x) \lel \delta (n) + l/2$.
   Moreover, $K(x|y) \lel l/2$ since we can retrieve $x$ by
providing $l/2$ bits.
  Therefore, $K(x|y)+K(y|x) \lel l/2+ \delta (n) + l/2$.
  Since the left-hand side has value at most $d$, the largest
$l$ we can choose is 
$l \eql d- \delta (n)$.

  This shows that the number $\#B_3(d,x,n)$ of $y$'s 
  such that $E_3 (x,y) \leq d$ satisfies
\[ \log \#B_3(d,x,n) \gel   \frac{\delta (n) +d}{2} . \]

\noindent($\leq$)
  Assume, to the contrary, that there are at least $2^{(d+\delta(n))/2
+c}$ elements $y$ of length $n$ such that $E_3(x,y)\le d$ holds, with
$c$ some large constant to be determined later.
  Then, for some $y$,
 \[
   K(y|x) \geq \frac{d+\delta(n)}{2} +c .
 \]
  By assumption, $K(x) \eql n- \delta (n),\ K(y) \lel n$.
  By the addition theorem \ref{e.addition} we find
 $ n + (d- \delta (n))/2+c \lel n + K(x|y)$.
  But this means that
 \[
  K(x|y) \gel \frac{d- \delta (n)}{2}+c ,
 \]
  and these two equations contradict $K(x|y)+K(y|x) \leq d$ for 
  large enough $c = O(\log n)$.
 \end{proof}

  It follows from our estimates that in every set of low Kolmogorov
complexity almost all elements are far away from each other in terms
of the distance $E_1$.

  If $S$ is a finite set of low complexity (like a finite
initial segment of a recursively enumarable set) then almost
all pairs of elements in the set have large information distance.
Let the Kolmogorov complexity $K(S)$ of a set be the length of a
shortest binary program that enumerates $S$ and then halts.

  \begin{formalsl}{Theorem}\label{topology.theorem}
  For a constant $c$, let $S$ be a set with $\# S=2^d$ and
 $K(S)=c\log d$.
  Almost all pairs of elements $x,y \in S$ have distance $E_1(x,y)
\geq d$, up to an additive logarithmic term.
 \end{formalsl}

  The proof of this theorem is easy.
  A similar statement can be proved for the distance of a string $x$
(possibly outside $S$) to the majority of elements $y$ in $S$.
  If $K(x) \geq n$, then for almost all $y \in S$ we have $E_1(x,y)
\geq n+d \pm O(\log dn)$.

\section*{Acknowledgment}
We thank 
John Tromp for many useful
comments and for shortening the proof of Theorem~\ref{t.excess},
 Zolt\'an F\"uredi for help with the proof of Lemma~\ref{lem.zf},
Nikolai K. Vereshchagin for his comments on maximum overlap
and minimum overlap in Section~\ref{s.conv}, 
and an anonymous reviewer for comments on Section~\ref{s.thermo}.

\end{document}